\documentclass{agujournal}

\journalname{JGR-Planets}

\usepackage[UKenglish]{babel} 					
\usepackage{amsmath} 							
\usepackage{amssymb} 							
\usepackage{fancyref} 							
\usepackage[font=small, labelfont=bf]{caption}	
\usepackage{subcaption}
\usepackage[super]{nth} 						
\usepackage{wrapfig}							
\usepackage{hyperref}							
\hypersetup{colorlinks=true}

\newcommand{\apj}{Astrophys. J.}
\newcommand{\apss}{Astrophys. Space Sci.}
\newcommand{\icarus}{Icarus}
\newcommand{\grl}{Geophys. Res. Lett.}
\newcommand{\jgr}{J. Geophys. Res.}
\newcommand{\jcp}{J. Chem. Phys.}
\newcommand{\planss}{Planet. Space Sci.}
\newcommand{\ssr}{Space Sci. Rev.}

\newcommand{\half}		{\tfrac{1}{2}}

\newcommand{\Rar}		{\Rightarrow}

\begin{document}

\title{Evidence for a Localised Source of the Argon in the Lunar Exosphere}

\authors{Jacob A. Kegerreis\affil{1},
		Vincent R. Eke\affil{1}, 
		Richard J. Massey\affil{1},
		Simon K. Beaumont\affil{2}, 
		Rick C. Elphic\affil{3}, and 
		Lu{\'i}s F. Teodoro\affil{4}}

\affiliation{1}{Institute for Computational Cosmology, Department of Physics, Durham University, South Road, Durham, DH1 3LE, UK}
\affiliation{2}{Centre for Sustainable Chemical Processes, Department of Chemistry, Durham University, South Road, Durham, DH1 3LE, UK}
\affiliation{3}{NASA Ames Research Center, Moffett Field, CA, USA}
\affiliation{4}{BAER/NASA Ames Research Center, Moffett Field, CA, USA}


\correspondingauthor{Jacob A. Kegerreis}{jacob.kegerreis@durham.ac.uk}


\begin{keypoints}
	\item We test various proposed explanations for observed features of the lunar argon exosphere.
	\item Explaining the ``bulge'' in argon density over the maria requires either a highly localised source or rapid turnover of argon.
	\item Seasonally varying cold traps could explain the long-term variation in the global argon density observed by LADEE.
\end{keypoints}

\begin{abstract}
We perform the first tests of various proposed explanations for  
observed features of the Moon's argon exosphere,
including models of: spatially varying surface interactions; 
a source that reflects the lunar near-surface potassium distribution;
and temporally varying cold trap areas.
Measurements from the Lunar Atmosphere and Dust Environment Explorer (LADEE) 
and the Lunar Atmosphere Composition Experiment (LACE) 
are used to test whether these models can reproduce the data.
The spatially varying surface interactions hypothesized in previous work cannot
reproduce the persistent argon enhancement observed over the western maria.
They also fail to match the observed local time of the near-sunrise peak 
in argon density, which is the same for the highland and mare regions,
and is well reproduced by simple surface interactions with a 
ubiquitous desorption energy of 28~kJ~mol$^{-1}$. 
A localised source can explain the observations, with a trade-off between
an unexpectedly localised source or an unexpectedly brief lifetime of argon
atoms in the exosphere. 
To match the observations, a point-like source requires source and loss
rates of $\sim$${1.9\times10^{21}}$~atoms~s$^{-1}$. 
A more diffuse source, weighted by the near-surface
potassium, requires much higher rates of
$\sim$${1.1\times10^{22}}$~atoms~s$^{-1}$, corresponding to a mean
lifetime of just 1.4~lunar~days. 
We do not address the mechanism for producing a localised source, but
demonstrate that this appears to be the only model that can reproduce the
observations. 
Large, seasonally varying cold traps could explain the long-term 
fluctuation in the global argon density observed by LADEE, but not that by LACE.
\end{abstract}

%



%
%
%

\section{Introduction}
\label{sec:intro}

The Moon possesses our nearest example of a surface-bounded
exosphere, the most common type of atmosphere in the solar system. 
As the atoms constituting an exosphere do not interact
with one another during their ballistic trajectories over the surface,
different species form independent systems. Their exospheric densities 
and variation with local time depend upon the sources, sinks, and
surface interactions for that particular species. Hence, studying the
lunar exosphere has the potential to teach us about the solar wind,
the lunar interior and outgassing, the efficiency of volatile
sequestration in polar cold traps, and the kinetics of adsorption and
desorption in low pressure environments \citep{Stern1999,Watson+1961,Wieler+Heber2003}.

Argon is a particularly well-studied species in the lunar exosphere,
having been first detected by the Lunar Atmosphere Composition 
Experiment (LACE), which measured the
$^{40}$Ar/$^{36}$Ar ratio at the surface to be approximately $10$
\citep{Hoffman+1973}. 
This implied that the more important source of argon was
radioactive decay of $^{40}$K to $^{40}$Ar, rather than solar-wind derived
$^{36}$Ar. The LACE results showed that the argon exospheric density
decreased through the night and had a rapid increase that began just before
sunrise, typical of a condensible gas that adsorbs to the cold
nighttime surface and desorbs at dawn \citep{Hodges+Johnson1968}. In addition to 
this daily variation, there was a longer term decrease by a factor of
$\sim$2 seen during the nine lunar days of
observations \citep{Hodges1975}. 

The Lunar Atmosphere and Dust Environment Explorer (LADEE) 
orbital mission produced a wealth of data
concerning the lunar exosphere at altitudes from 3--140~km \citep{Elphic+2014}. 
As well as measuring the daily and long-term variations in argon density
during its 5 month mission, the Neutral Mass Spectrometer
\citep[NMS,][]{Mahaffy+2014} also determined the vertical 
structure of the exosphere and the variation with selenographic
longitude. This led to the discovery that there was an
enhancement in the argon exospheric density over the western
maria, dubbed the argon ``bulge'' by \citet{Benna+2015}. 
The long-term variation in the argon
abundance was $\sim$28\% during the LADEE mission,
much smaller than had been seen 40 years earlier by LACE over similar time periods.
However, \citet{Hodges+Mahaffy2016}
noted that ``the absence of sensitivity-related tests of the Apollo 17
instrument allows the possibility that the 1973 results were in part
artefacts.''

Different models of aspects of the lunar $^{40}$Ar system have been
created to help interpret the available data, both in terms of the
outgassing rate from the surface and the corresponding sinks that
are necessary to yield the measured exospheric
density. \citet{Hodges1975} used the LACE data and Monte Carlo methods 
to simulate an argon exosphere to constrain both the source
rate and the surface interactions.
\citet{Grava+2015} also employed a Monte Carlo technique to follow
an initial injection of argon atoms through their lifetimes in the
exosphere, concluding that approximately 10\% of the area of permanently
shadowed regions \cite[PSRs,][]{Mazarico+2011} is needed to cold trap atoms in order to 
provide a sufficiently high loss rate to match the LACE long-term decline in argon
exospheric density. If a continuous background source had been included in their
model, then larger cold traps would have been required to deplete the
exospheric argon density rapidly enough. 

Using their model, \citet{Grava+2015} suggested that long-term
variations in the exospheric density can be ascribed to sporadic
moonquakes. \citet{Benna+2015} noted the possibility of tidal stress as the
source of the LADEE variation.
In contrast, \citet{Hodges+Mahaffy2016} proposed that
seasonal fluctuations in the total cold trap area are responsible for the 
smooth, mission-long variations in argon density measured by LADEE
\citep{Benna+2015}. 

More than one proposed explanation also exists for the bulge -- the 
persistent enhancement of exospheric argon localised
over the western maria. \citet{Benna+2015} noted the similarity
between the longitudinal variation in argon and the map of
near-surface potassium returned by the Lunar Prospector Gamma Ray
Spectrometer \citep[LPGRS,][]{Lawrence+1998}, suggesting a localised source. 
However, \citet{Hodges+Mahaffy2016} asserted that 
the lifetimes of argon atoms in the lunar exosphere are too long
for them to reflect their source locations. Instead, they suggested that 
the bulge results from lower desorption energies at these longitudes, 
which would cause argon to spend less time residing on the surface where it
cannot be measured.

In this paper, we develop a new model of the lunar argon exosphere using
Monte Carlo methods. This approach is similar to those described by
\citet{Smith+1978}, \citet{Hodges1980a}, and \citet{Butler1997}, in
their studies of the helium and water exospheres of the Moon and
Mercury. We apply our algorithm to address the questions of which 
-- if any -- of the proposed models could be
responsible for the longitudinal and long-term variations in the 
argon densities measured by LADEE. Specifically, we produce the first
simulations with: spatially varying surface interactions; a source that
reflects the lunar near-surface potassium distribution; and
seasonally varying cold trap areas.

Section~\ref{sec:data} contains a description of the data set used and
an overview of the different aspects of the lunar argon exosphere that the
LADEE data can constrain. Our model is outlined in
section~\ref{sec:model}, and the results and
their implications for the source, sinks and regolith interactions of
argon atoms are described in section~\ref{sec:results}.

\section{Data}	
\label{sec:data}

The NMS on LADEE measured the density of argon (and other
species) in the lunar exosphere from \nth{22}~November~2013 to
\nth{17}~April~2014 at a wide range of altitudes, longitudes, and
local times of day at latitudes within $30^\circ$ of the equator.  
Derived data, including background-subtracted argon number densities at altitude, 
were obtained from The Planetary Atmospheres
Node of NASA's Planetary Data System.

We apply two cuts to the entire LADEE argon data set, to produce the
subset of data used here. In the full data set, any
densities that are negative after the background subtraction (due to noise) 
have had their values set to zero, 
which causes the mean to be artificially high if these are either 
included as zeros or discarded. We only use data in bins of local time of day 
and selenographical longitude for which more than half of the observations are
positive. In this way, the medians of the resulting sets of measurements
should not be biased by this prior treatment of negative values.
Also, an unaccounted-for temperature dependence
of the instrument background can affect the densities just
after midnight (M. Benna, personal communication, 2016). This
problem is only important at this local time of day and for very low
densities, where the instrument background was large compared with the
signal. We therefore discard data at local times of day 180$^\circ$--$265^\circ$ 
(where 0$^\circ$ is noon, 90$^\circ$ is sunset, 180$^\circ$ is midnight, 
and 270$^\circ$ is sunrise). Fortunately, the
LACE data largely fill the overnight gap, and we
supplement the LADEE measurements using the results presented by
\citet{Hodges1975}.

Four complementary aspects of the argon exosphere can readily be studied:
(1) the change in density with altitude; (2) the long-term variation
in the global density during the months of LADEE's operation; (3) the
density distribution with local time of day; and
(4) the dependence on selenographical longitude, 
showing the bulge over the western maria \citep{Benna+2015}. 

\subsection{Densities at Altitude} 
\label{sec:data:altitude}	

In order to study the final three of these distributions, 
we need to account for the measurement altitude varying from
3 to 140~km. Following convention, and for comparisons with the LACE data, 
we convert all measurements to
the corresponding densities that would have been measured at the surface.
To do this, we consider the expression derived by
\citet{Chamberlain1963} linking the number density in an exosphere as a function of
height, $n(h)$, to the number density and temperature at the surface,
$n_0$ and $T$ respectively. For a spherical body with mass $M$ and
radius $r$, 
\begin{linenomath*}
	\begin{equation}
	n(h) = n_0\,\exp\left[- \dfrac{GMm}{kT} \left(\dfrac{1}{r} - \dfrac{1}{r+h}\right)\right] \;,
	\label{eqn:Chamberlain}
	\end{equation}
\end{linenomath*}
where $m$ is the mass of the particle, $G$ is the gravitational constant, 
and $k$ is the Boltzmann constant. This is the generalised form of the (isothermal) barometric law. 
We refer to this as a ``Chamberlain distribution''. The altitude dependence used by
\citet{Benna+2015}, which results from assuming a 
constant acceleration due to gravity, is the first-order expansion
of Eqn.~(\ref{eqn:Chamberlain}) for small $h$. 

The surface temperature varies as a function of latitude and local
time, with a particularly rapid variation around the
terminators. Lateral transport of molecules implies that the density
at altitude will not reflect only the single sub-detector surface
temperature, because particles reaching the detector will have originated at 
a variety of locations and temperatures. 
Therefore, we expect the real distribution to be a sum of many 
Chamberlain distributions for different temperatures, weighted by the number 
of particles that come from each one. As a practical model,
we approximate this distribution with a sum of just two Chamberlain distributions
at different temperatures and find the best-fit parameters 
at all times of day using our simulations, 
as detailed in section.~\ref{app:model:altitude}.

\begin{figure}[t!]
	\centering	
	\includegraphics[width=\columnwidth]{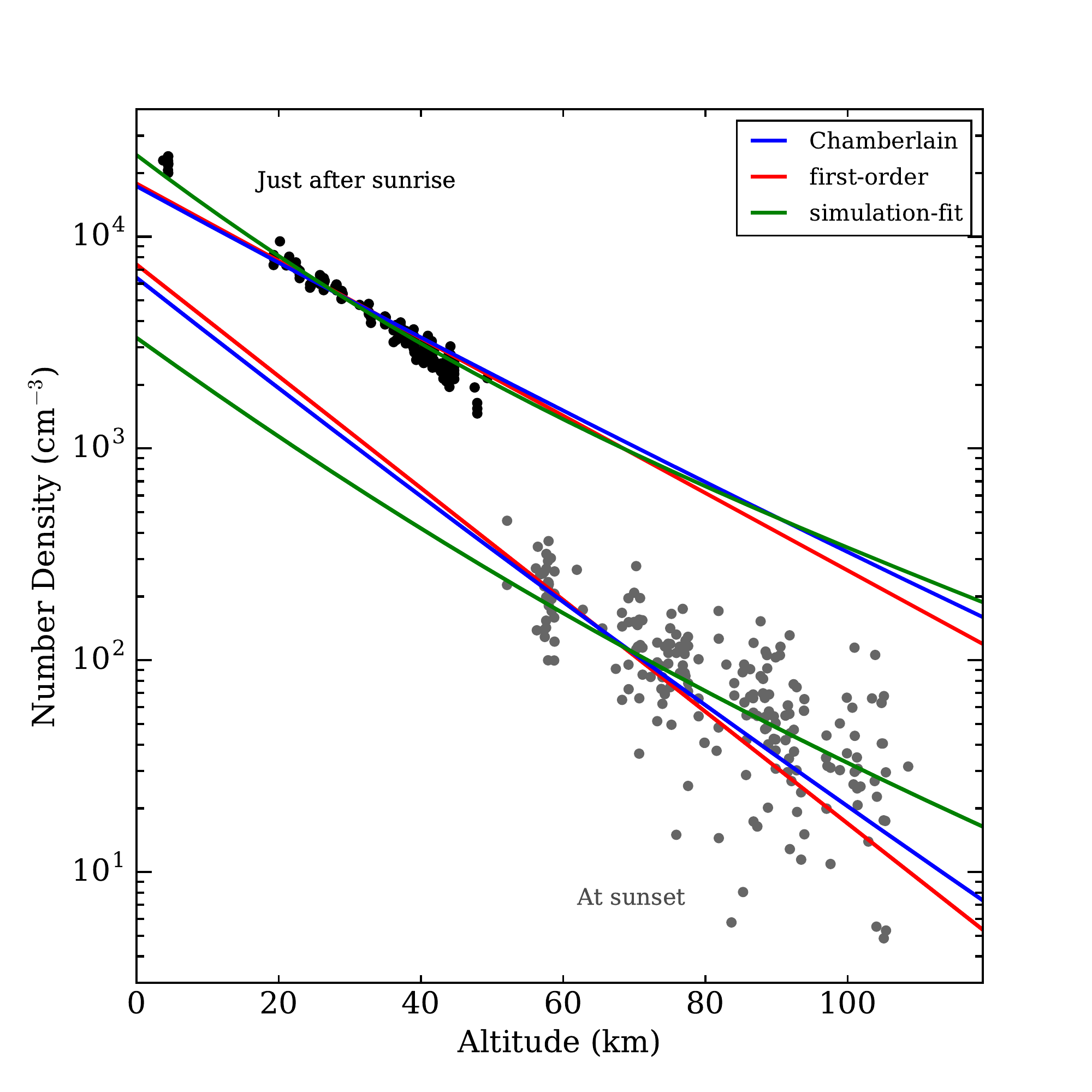}
	\caption{Two examples of the change of density with altitude near the terminators. 
		The black and grey points show the LADEE data just after 
		sunrise (273$^\circ$) over the maria and at sunset (90$^\circ$) 
		over the highlands respectively.
		The coloured lines show the three models at the same local times
		of day, scaled in magnitude to match the data. The data have had the 
		long-term variation extracted, as described in section~\ref{sec:data:long_term}.
		\label{fig:altitude_examples}}
\end{figure}

Fig.~\ref{fig:altitude_examples} shows two illustrative examples of the altitude
variation of the LADEE data near the terminators, 
where the three models described above differ most from each other.
It is apparent that the choice of extrapolation to zero altitude can significantly
affect the inferred number density at the surface.
\citet{Hurley+2016} found a similar discrepancy between a Chamberlain profile
and their model of the helium exosphere.

The LADEE data cover only a small range of altitudes at most 
local times of day. So, we use our simulations to test how accurately the 
three different models predict the simulated density at the surface
at every time of day, from observations taken at the average LADEE altitude of 60~km. 
We found that our ``simulation-fit'' model of two Chamberlain 
distributions with different temperatures successfully predicts 
the density at the surface to within 12\% everywhere. 
This compares with overestimates as high as 
337\% and 416\% for the single Chamberlain
distribution and its first-order expansion respectively, 
which both use single temperatures (from the \citet{Hurley+2015} 
model described in section~3.4). 
These deviations are most pronounced near the terminators.
For example, just before sunrise, where the subdetector 
surface temperature is very low, many particles will also be detected that 
originated at the hot surface after sunrise, which the Chamberlain and first-order
models cannot account for. 
Away from the terminators, in regions where the surface temperature varies only
slowly with local time of day, all three extrapolations 
correctly predict the simulation's density at the surface to within a few per cent, 
although the first-order model does less well at higher altitudes.
We use the simulation-fit model to infer all
LADEE argon densities at the surface reported in this paper.

The agreement of the simulation results with the LADEE altitude data 
at all local times of day is also 
good evidence that our underlying models for the simple thermal 
desorption of argon atoms from the surface are appropriate. 
The data also do not show any clear differences between the altitude
distributions above the mare and highland regions.

\subsection{Long-Term Variation} 
\label{sec:data:long_term}

\begin{figure}[t!]
	\centering	
	
	\includegraphics[width=\columnwidth]{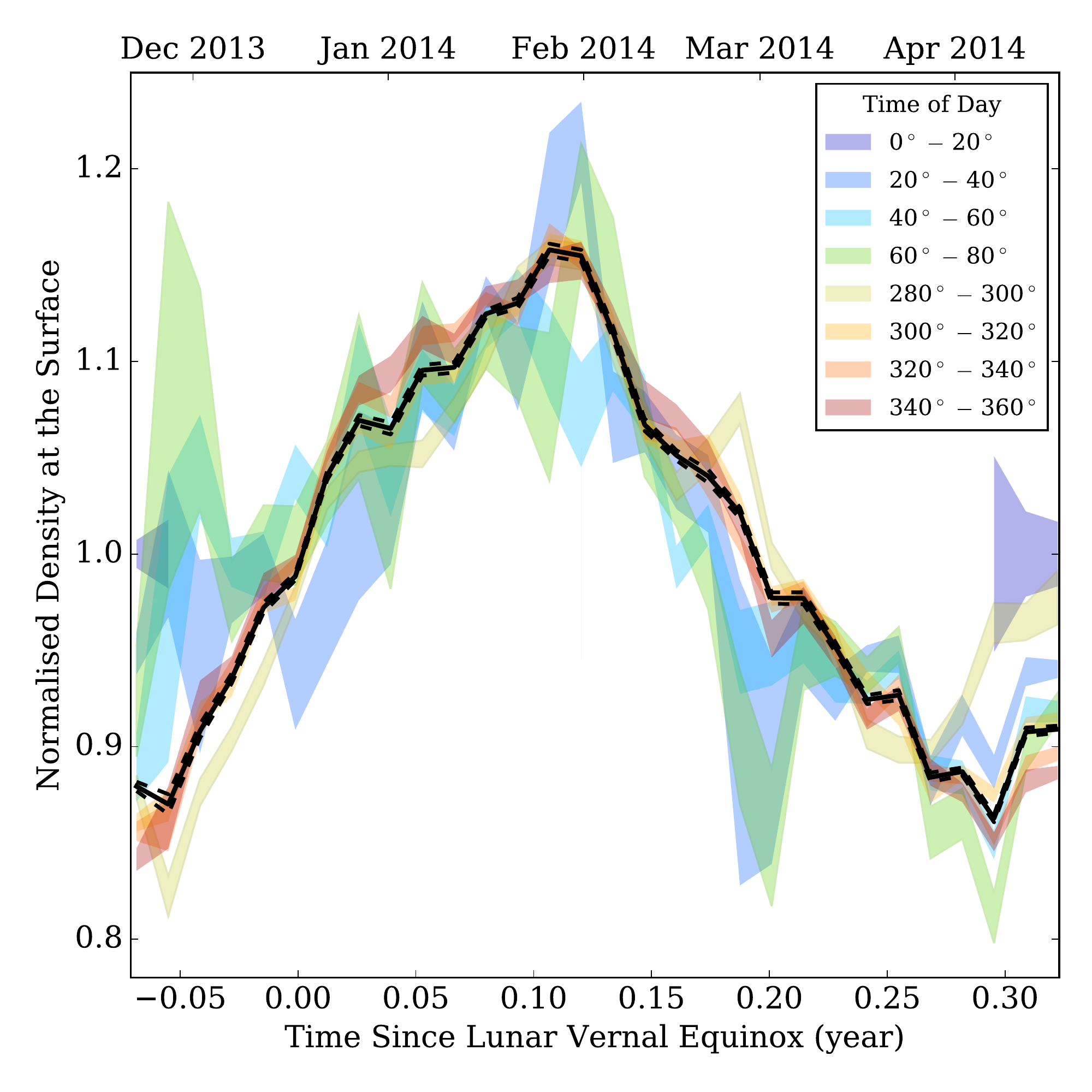}
	
	\caption{The long-term variation in the LADEE argon density, normalised by the
		mean density. The different colours show the median variation and its
		1$\sigma$ uncertainty at the
		corresponding local time of day, as given in the legend. The solid black
		line shows the total mean variation, weighted by the errors on the
		time-of-day values. Dashed black lines represent the 1$\sigma$ uncertainty on
		this mean. Time is measured from the lunar vernal equinox on 
		17 December \citep{Archinal+2011}.
		\label{fig:long_term_data}}
\end{figure}

As the argon abundance varies dramatically with the local time of day, to
determine the long-term variation we first split the LADEE data into bins of
20$^\circ$ in local time and calculate the mean density for each bin.
The long-term variation is then calculated by subtracting the corresponding 
mean value from every measurement. 
To combine the data across all local times, we then divide by the 
same mean value to give the normalised deviation at a given long-term time.
The median deviations are shown in Fig.~\ref{fig:long_term_data}, 
along with the weighted mean of all of these curves. 
The general agreement across the different times of day is notable,
as is the relatively smooth variation. 

The peak-to-peak change in density is 28\%.
\citet{Benna+2015} noted the somewhat similar magnitude
and timescale of the variation to that observed
by LACE, and offered that transient changes in the
release rate of argon from the Moon's interior are a plausible
cause of this variability. \citet{Hodges+Mahaffy2016} instead suggested that it
is part of a periodic fluctuation with a period of
half a year. They proposed that this is driven by seasonal
variations in the polar cold trap areas.

All subsequent figures in this section show densities at the
surface with the mean long-term variation removed from the data. 
This is done by dividing each data point by the normalised long-term variation
at the time of the measurement. This reveals what LADEE would have 
observed had the exosphere been in a steady state. 

\subsection{Local Time of Day} 
\label{sec:data:local_time}

\begin{figure}[t!]
	\centering	
	\includegraphics[width=\columnwidth]{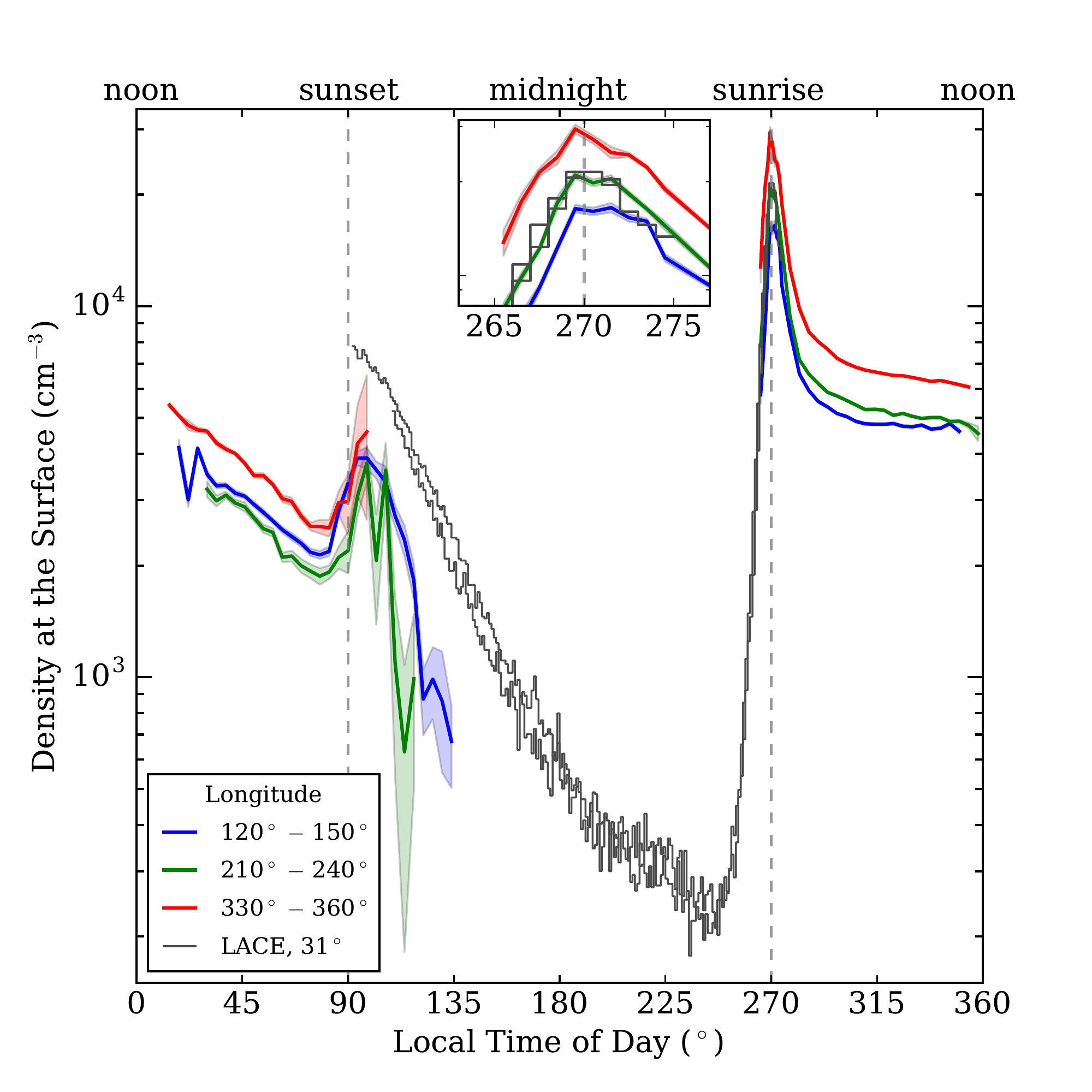}
	\caption{The variation of the argon density with local time of
		day. The coloured lines are typical examples of the LADEE data 
		across the mare (red) and highland (blue and green) selenographic longitude regions, 
		as given in the legend.
		The shaded areas represent the $\pm 1 \sigma$ uncertainties.
		The inset plot zooms in on the region around the sunrise peak. 
		The grey lines are two sets of LACE data
		separated by a few lunar days \citep{Hodges1975} and
		normalised to match the density from LADEE at LACE's location
		at sunrise. 
		A vertical slice would yield the
		distribution of density with longitude at that 
		time of day, as shown in Fig.~\ref{fig:bulge_data}. 
		\label{fig:local_time_data}}
\end{figure}

Fig.~\ref{fig:local_time_data} shows the distribution of argon with
local time of day from LADEE and LACE data, with the LADEE
data corrected for altitude and long-term variability as described
above. The figure shows just three representative examples
across the mare and highland regions for clarity. 
The LADEE (and LACE) data show very similar behaviour at all selenographic
longitudes. In particular, the timing of the sunrise peak is insensitive to
the longitude, as highlighted by the inset panel, occurring
at local times between ${269^\circ\pm1^\circ}$ and ${272^\circ\pm1^\circ}$. 
However, the argon density is greater in the
maria than that in the highlands by a factor that ranges from a few tens of
per cent up to almost a factor of 2 at sunrise.
The LACE densities have been rescaled in amplitude to match the
inferred surface density at sunrise from LADEE at the location of LACE, 
to extract the long-term variation between the data set. 
Note that the lowest late-night LACE measurements may be below the instrument's
sensitivity \citep{Hoffman+1973}.

The large peaks in density around sunset and sunrise are both 
fed by particles migrating away from noon, where the higher 
temperatures mean larger hops. At sunset, the temperature is lower, 
so particles do not hop very far, but it is not yet cold enough to
trap argon on the surface for long periods of time. 
Any particles that do stick to the surface at night will rotate
with the Moon toward sunrise, creating the enormous peak when they 
warm up at dawn and reenter the exosphere. 
This peak extends back into the night because the particles 
fly in all directions, with a typical hop distance of a 
few degrees at post-sunrise temperatures.

\subsection{Selenographic Longitude -- The Bulge} 
\label{sec:data:bulge}

\begin{figure}[t!]
	\centering	
	\includegraphics[width=\columnwidth]{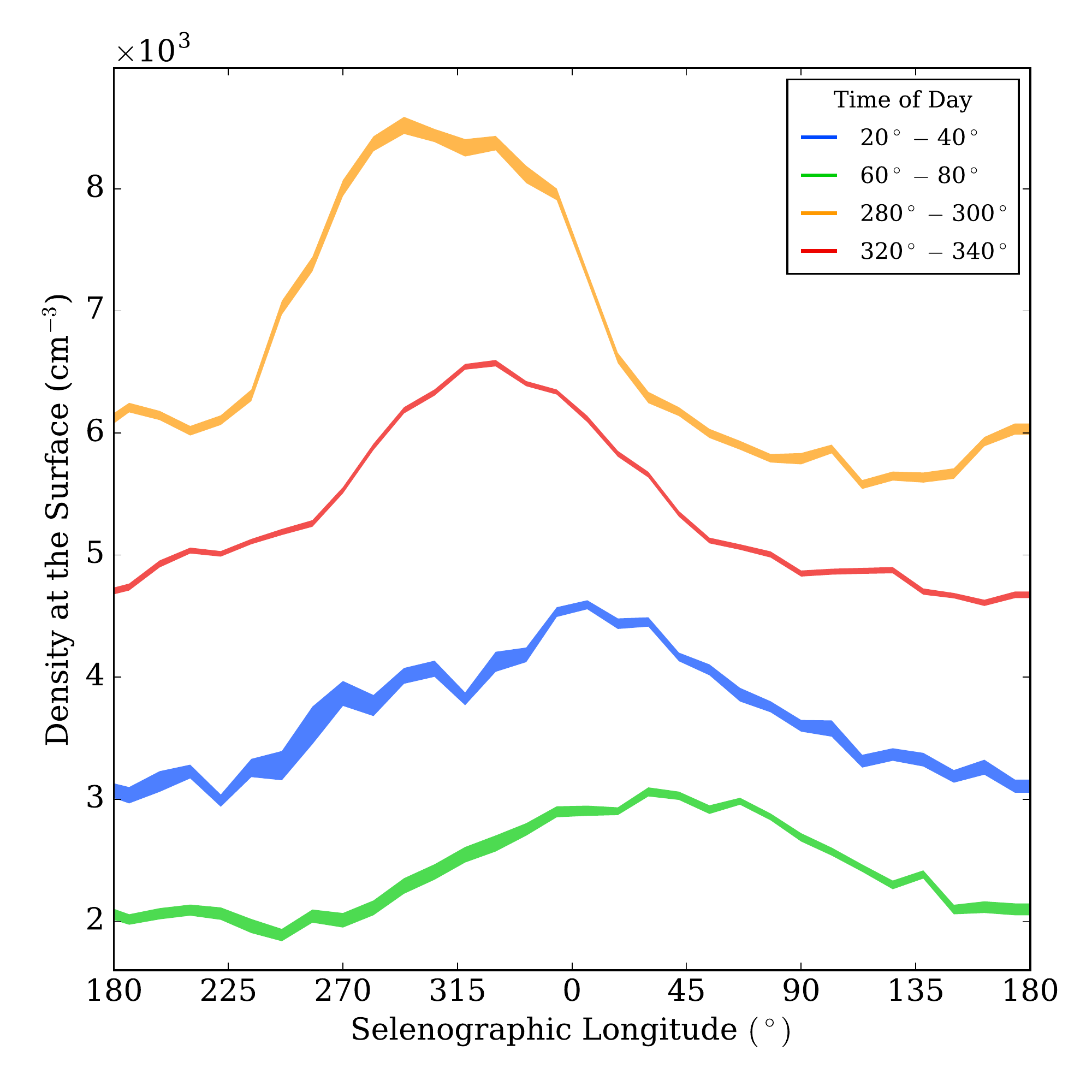}
	\caption{The variation of argon density with selenographic
		longitude from LADEE, showing the bulge over the western maria and
		its evolution through the lunar day. The
		coloured regions show a selection of different local times of day, 
		the shaded areas represent the $\pm 1 \sigma$ uncertainties.
		A vertical slice would yield the
		distribution of density with time of day at that longitude, as
		shown in Fig.~\ref{fig:local_time_data}. 
		\label{fig:bulge_data}}
\end{figure}

The change of density with selenographic longitude is shown in
Fig.~\ref{fig:bulge_data} for different slices in local time of day. As was
evident from Fig.~\ref{fig:local_time_data}, the density at all local times of
day is highest at some point over the maria (longitudes from
270$^\circ$--$45^\circ$). The peak near sunrise is located
over the western maria in the region of the Procellarum KREEP Terrain (PKT),
which is rich in $^{40}$Ar's parent, $^{40}$K \citep{Jolliff+2000}.
Along the different curves of fixed local time, 
the selenographic longitude of the peak argon density drifts
systematically from $\sim$300$^\circ$ at sunrise to $\sim$45$^\circ$ at
sunset.

These distributions all reflect a complex interplay between
sources, sinks and, most importantly, surface interactions of the
argon atoms. Consequently, the variation of argon density with both
selenographic longitude and local time of day offer the opportunity to
distinguish between different models for the argon exosphere.

\section{Model}	
\label{sec:model}

In this section, we describe the main processes and input parameters
in our model. Appx.~\ref{app:model} contains full 
details of aspects of our model that differ from similar previous
studies such as that by \citet{Butler1997}. The simulation code itself
is publicly available with documentation at
\href{http://www.icc.dur.ac.uk/index.php?content=Research/Topics/O13}{icc.dur.ac.uk/index.php?content=Research/Topics/O13}.

The central idea is to follow one particle at a time throughout 
its life, then repeat this for many particles to build up a 
model of the lunar argon exosphere.  
Each simulation particle represents a number of argon atoms, and
in between its creation and eventual loss from the system, it
migrates in a series of interactions with the surface and ballistic
hops. The models for each of these various processes are now described in
turn.

\subsection{Source} 
\label{sec:model:source}

Most of our simulations are for a steady-state exosphere in which the
continuous source rate matches the loss rate, after an initialisation period 
in which more particles are created than are lost. 
We assume a continuous source given the relatively smooth variation observed by 
LADEE, which suggests a lack of dramatic transient source events. 
The source rate, the mean lifetime,
and the total number of particles in the equilibrium system are directly
related -- knowing any two determines the third. In our simulations, 
we can investigate a range of mean lifetimes by varying the sinks.
The total amount of argon in the exosphere is then scaled to match that 
from the LADEE measurements by setting the source rate. 

The value for the source rate is implicitly varied in a range that
reflects the uncertainty in the amount of potassium in the Moon and the
effectiveness with which radiogenic $^{40}$Ar reaches the surface.
\citet{Killen2002} modelled argon's production and diffusion from the
potassium in the crust and estimated that argon enters
the exosphere at a rate in the range of
{3.8--5.5$\times10^{20}$}~atoms~s$^{-1}$. These values correspond to
only a few per cent of the $^{40}$Ar that is created inside the Moon 
\citep{Hodges1975}, so we also investigate
significantly higher source rates.

Following the discussion from \citet{Benna+2015}, 
we wish to test if a (continuous) localised source can reproduce the
argon bulge over the western maria and, if so, what source and loss rates 
this would require. Thus, we 
use either a global source, where the argon particles
appear isotropically at random locations on the spherical surface, or a local
source, where they appear preferentially at locations with higher
near-surface potassium concentrations. As noted by \citet{Benna+2015}, while argon is
expected to originate from deep, molten sources, there may be preferred
diffusion pathways up through the same region marked by the potassium
and PKT. The LPGRS, and more recently gamma-ray spectrometers on board
Chang'E-1 and Chang'E-2, measured the potassium
abundance and distribution in the top metre or so of the regolith
\citep{Prettyman+2006,Zhu+2011,Zhu+2015}. 
For the localised source model, we use the LPGRS 
potassium map to weight the source distribution for the
simulation particles, such that the probability of being sourced at a given
location is proportional to the local potassium concentration.

\subsection{Sinks}
\label{sec:model:losses}

There are two main ways in which particles can be lost from our
simulations: interactions with photons or charged particles from the
Sun; and cold trapping on the surface in the permanently shadowed polar regions. 
Our implementation of these physical processes is described in the
following subsections. We include the possibility of gravitational
escape in our simulations, but for argon this has a negligible effect.

\subsubsection{Solar Radiation}

A particle in flight on the dayside may be lost due to a variety of
processes, the most important of which are photoionisation and
charge-exchange with solar wind protons \citep{Grava+2015}. An 
ionised particle will rapidly be driven either away from or
into the Moon's surface by the local electromagnetic field. 
Those that impact the surface may be neutralised and ``recycled'' back into the 
exosphere. 
Following \citet{Butler1997} and \citet{Grava+2015}, 
such processes can be combined to give a single
solar radiation destruction timescale, $\tau$.
The probability of loss during a flight of time $t$ is then 
\begin{linenomath*}
	\begin{equation}
	P(t) = 1-e^{-t/\tau} \;.
	\end{equation}
\end{linenomath*}
For each particle hop, our algorithm picks a random number from a
uniform distribution between 0 and 1. If this lies below $P(t)$, then
the particle is removed from the simulation.

During LADEE's operation, the mean solar wind speed and proton density 
were 400~km~s$^{-1}$ and 5~cm$^{-3}$ respectively
(from the GSFC/SPDF OMNIWeb database interface at 
\href{http://omni-web.gsfc.nasa.gov}{omni-web.gsfc.nasa.gov}), 
giving a proton flux of $2 \times 10^{8}$~cm$^{-2}$~s$^{-1}$.
Multiplying this by the interaction cross section, $2 \times 10^{-15}$~cm$^{2}$ \citep{Nakai+1987}, 
gives a rate for proton-argon charge-exchange of $4 \times 10^{-7}$~s$^{-1}$. 
Adding the photoionisation rate from \citet{Huebner+1992}, 
$3 \times 10^{-7}$~s$^{-1}$, taking the inverse, 
and finally dividing by the recycling fraction of 0.5 \citep{Poppe+2013}, 
gives a timescale of $\tau=3 \times 10^{6}$~s.
Including the recycling process by simply increasing the timescale 
is analogous to assuming that a recycled ion reenters the exosphere
without travelling a long time or distance.
Many of these values have significant uncertainties,
such as the cross section which has a somewhat broad peak around 1~keV protons, 
and the solar wind speed and proton density, which fluctuate significantly.
Thus, this error on this timescale is likely at least a factor of 2.

\subsubsection{Cold Traps}

If a particle lands in a permanent cold trap near the poles, then it
is assumed to stick there indefinitely. Like previous models, we adopt
a stochastic approach. When a particle lands
within $15^\circ$ of the north or south pole, it has a probability of
being trapped, given by the total fractional area of cold traps in
that polar region. Various estimates have been made for the size and
distribution of cold traps. The appropriate cold trap area depends on the
surface interaction of the specific species and the complicated
processes that may occur after landing in a cold trap, either to secure
or remove the particles \citep{Schorghofer+Aharonson2014,Chaufray+2009}.
As the uncertainty on the effective cold trap area is large, we leave this as a
free parameter. 

\begin{figure}[t!]
	\centering	
	\includegraphics[width=\columnwidth]{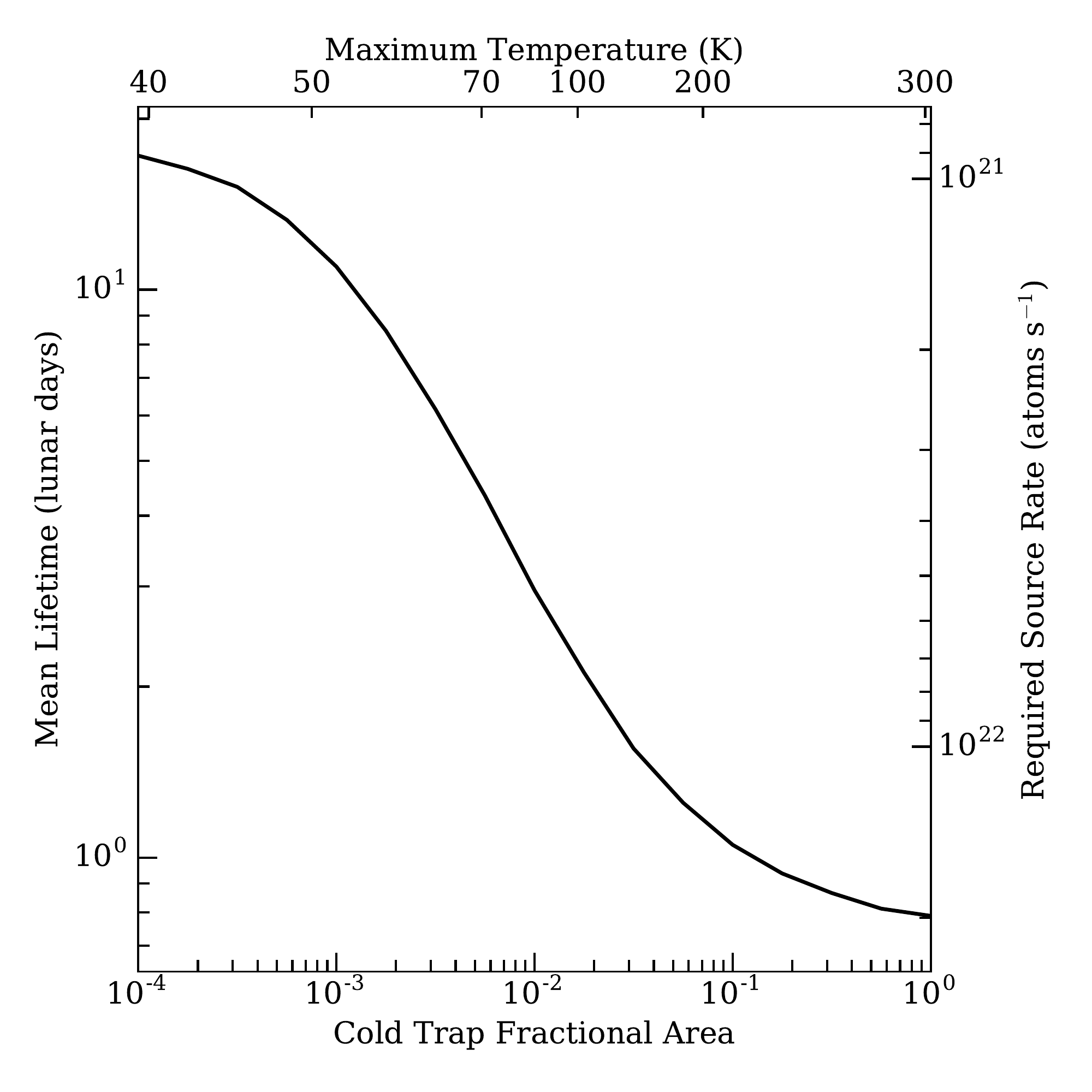}
	\caption{The mean lifetime of the simulated argon particles for
		different cold trap areas as fractions of the polar
		15$^\circ$. The right axis shows the source rate of argon that would
		be required to match the inferred LADEE total argon abundance. The top axis
		gives the maximum surface temperatures from Diviner that correspond to
		the assumed areas covered by cold traps.
		\label{fig:lifetime_from_cold_traps}}
\end{figure}

Fig.~\ref{fig:lifetime_from_cold_traps} shows the
mean lifetime (and corresponding source rate) of argon atoms in the exospheric
system for different cold trap areas, 
keeping the photodestruction timescale constant and using the
surface interaction models described in the following subsection. Larger
cold traps result in shorter lifetimes, and a larger source rate is then
required to maintain the same total number of argon atoms. 
Also shown in Fig.~\ref{fig:lifetime_from_cold_traps}, for reference, 
are some maximum surface temperatures and their corresponding cold trap areas, 
as inferred from Diviner data \citep{Vasavada+2012}. 
This is done by relating a given cold trap area to the same-size area 
that never exceeds a certain temperature.

For our default surface interaction model, the total
argon content in the simulation is set at 
${4\times10^{28}}$ atoms to match the LADEE abundance measurements. 
The main uncertainty in this number arises from the long-term and 
selenographic variations in density measured by
LADEE, amounting to about 44\%.

\subsubsection{Seasonal Cold Traps}
\label{sec:model:seasonal}

We include the possibility of seasonal cold traps in our model 
in addition to the permanent cold traps.
These have fractional areas in the north
and south polar 15$^\circ$ that grow and shrink periodically as
\begin{linenomath*}
	\begin{equation}
	\begin{split}
	f_{\text{N}} &= {\rm max}\left[0\,,\,f_{\text{peak}} \sin\left(2\pi\,(t_{\text{yr}} - 0.5) \right)\right] \\
	f_{\text{S}} &= {\rm max}\left[0\,,\,f_{\text{peak}} \sin\left(2\pi \,t_{\text{yr}} \right)\right] \;,
	\label{eqn:seasonal_trap_sizes} 
	\end{split}
	\end{equation}
\end{linenomath*}
where $f_{\text{peak}}$ is the peak fractional area and $t_{\text{yr}}$ is the
time in units of years. Thus, the year begins with the vernal 
equinox and the southern trap reaches its maximum size one quarter of
the way through the year, followed by the autumnal equinox and the
northern peak in turn. The seasonal traps disappear completely in the 
summer half of the year for that pole. 
This asymmetry is required to match the half-year period suggested by the data 
(see Fig.~\ref{fig:long_term_data}).
If instead the seasonal traps varied symmetrically in both halves of the year, 
for example, sinusoidally as in Eqn.~\ref{eqn:seasonal_trap_sizes} but without truncation, 
then the summed seasonal trap area of both poles would 
vary with a period of 1 year -- or be constant if the maximum area at each
pole were the same.  
Thus, a model must have the same overall half-year period
to have the potential to explain the data.
This requires asymmetrical variation in summer and winter,
as is modelled simply by Eqn.~\ref{eqn:seasonal_trap_sizes}.

At the time a particle lands in a seasonal trap, 
there is a distribution of times at which 
the present cold traps first appeared (and hence the times at which
they will disappear), following Eqn.~\ref{eqn:seasonal_trap_sizes}.
Some will have appeared at the beginning of that pole's winter and others
potentially just before the particle landed.
A particle that lands in a seasonal cold trap
is released back into the exosphere at a randomly chosen time, 
following this distribution.

\subsection{Hop Trajectory} 
\label{sec:model:bounce}

In the absence of forces other than the gravity from the Moon, the complete paths of
particles -- including the landing position, time of flight, and
position and velocity at any altitude -- can be calculated
analytically from the starting location and velocity using Kepler's
laws (see Appx.~\ref{app:trajectory}). 
This approach is much less computationally intensive than integrating the paths numerically, 
even when positions and velocities are also calculated at altitude. 
This is how we transport particles in our simulation and, apart
from the time of flight, it matches the approach of
\citet{Butler1997} and \citet{Crider+Vondrak2000} for finding the landing position. 
For their time of flight calculation, 
they effectively assumed that the surface is
flat, leading to slightly underestimated flight times. 

For the initial velocity, the particles are assumed to have
accommodated to the surface temperature at their starting location and
to have a Maxwell-Boltzmann distribution in the exosphere for that
temperature. This means that the initial speed for a particle
leaving the surface must be drawn from the Maxwell-Boltzmann flux
distribution \citep{Brinkmann1970,Smith+1978}. As each particle
represents a small packet of argon atoms moving through the surface,
the random emission direction in the outward hemisphere needs to be
weighted by the component of the speed in the vertical
direction. The good fit of the simulation densities at altitude to the LADEE data
suggests that this model is appropriate. Another difference between our 
treatment and those of \citet{Butler1997} and \citet{Crider+Vondrak2000} 
is that they chose emission angles away from vertical, $\alpha$, 
from a non-isotropic distribution that was uniform between 0 and $\pi/2$, 
without including the $\sin(\alpha)$ term that accounts for the full 
area of the emission hemisphere.

\subsection{Surface Interaction} 
\label{sec:model:surface interaction}

The interaction of argon with the surface determines many of the
exosphere's characteristics. However, it is a complicated and
poorly understood process. 
In our model, once a particle lands it is assumed to adsorb
immediately. It will then reside upon the surface for 
some amount of time before being released. 
The residence time and the kinematics of the desorbed particle 
both depend sensitively upon the temperature at the location of the particle.
During the lunar night, the simulated particle may also ``squirrel'' down
into the regolith, to resurface at some time the following
day. This process turns out to be required to match the observed 
density distribution with local time of day while maintaining realistic 
desorption energy values, as is discussed in section~\ref{sec:results:time of day}.

The Diviner radiometer on the Lunar Reconnaissance Orbiter (LRO) 
produced temperature maps of the
lunar surface \citep{Vasavada+2012}. We use the analytical fit to the
Diviner data from \citet{Hurley+2015} to make a map of temperature as a
function of local time and location in square-degree bins. Following
\citet{Hurley+2015}, we also introduce a longitudinal Gaussian scatter with
${\sigma=4.5^\circ}$ into this map to account, statistically, for topographical relief.
This represents the temperature effects of, for example, 
the orientation of slopes near sunrise, which will receive sunlight at
different incidence angles, 
or the positions of ridges or craters that could see sunrise earlier or later respectively.

\subsubsection{Residence Time and Desorption} 

Every ${\Delta t=5}$~s (a time step is only introduced 
in this part of the simulation), the temperature is
re-calculated for a particle residing on the surface, 
to account for the Moon's rotation. This allows a residence
time, $t_{\text{res}}$, to be found using a standard modified Arrhenius equation
\citep{Bernatowicz+Podosek1991}: 
\begin{linenomath*}
	\begin{equation}
	t_{\text{res}}=\dfrac{h}{kT}\exp\left[\dfrac{Q}{RT}\right] \;, \label{eqn:t_res}
	\end{equation}
\end{linenomath*}
where $h$ is Planck's constant, $R$ is
the gas constant, $T$ is the temperature, and $Q$ is the desorption energy.

The probability of the particle desorbing from the surface in a given
time step is
\begin{linenomath*}
	\begin{equation}
	P(t_{\text{res}}) = e^{-\Delta t/t_{\text{res}}} \;.
	\end{equation}
\end{linenomath*}
If a uniform random number between 0 and 1 exceeds $P$, then the
particle is released and 
hops again. Otherwise, the simulation time is advanced by $\Delta t$
and the particle's position and the local temperature are
updated. This continues until the particle is released.  

For argon, a barrier-free adsorption process is expected, so heats of adsorption and 
activation energies for desorption can be equated, both corresponding to $Q$.
Experiments with argon on non-lunar aluminosilicates and mineral oxides
have shown that the desorption energy is typically around 8--10~kJ~mol$^{-1}$ \citep{Matsuhashi+Arata2001}. 
This increases slightly for more Lewis acidic materials,
although higher values were obtained for the heat of adsorption, 
up to 24~kJ~mol$^{-1}$ for some
acidified mineral oxides, possibly being representative of lower coverages or
corresponding to a small fraction of more strongly adsorbing sites.
The surface composition of the Moon is dominated by anorthosite, 
comprised primarily of a variety of silicate minerals \citep{Cheek+2013,Wieczorek+2006}.
Therefore, these terrestrial experiments provide a 
reasonable basis for estimating plausible values of $Q$.

On a low-energy metal oxide surface facet, \citet{Dohnalek+2002}
calculated the coverage-dependent desorption energy for argon
from temperature-programmed desorption (TPD) data and found it to 
increase from 8~kJ~mol$^{-1}$ to around 13~kJ~mol$^{-1}$ at very low coverages
(where only the highest energy sites should be occupied by argon atoms). 
If surface diffusion occurs readily and the coverage is low, 
then these strong adsorption sites may
be accessible to all adsorbing argon atoms. 
Direct calorimetric heats of adsorption on another
porous silica yielded $Q$ values of 18~kJ~mol$^{-1}$ 
\citep[again at low surface coverages of argon,][]{Dunne+1996}.
Interactions with the pristine lunar regolith may be even stronger \citep{Farrell+2015}, and
\citet{Bernatowicz+Podosek1991} found with a freshly crushed lunar sample that 
somewhat-higher energies of up to 31~kJ~mol$^{-1}$ (7.4~kcal~mol$^{-1}$) are plausible.
We conclude that experiments show we should expect argon-regolith interactions 
to involve energies around 10-30~kJ~mol$^{-1}$. 
However, until more in-situ experiments are performed, 
the precise value must be estimated empirically using 
observations of the exosphere as a whole.

\subsubsection{Squirrelling}

Argon particles enter the exosphere by migrating up through the porous
regolith from the Moon's interior \citep{Killen2002}. We propose that
some proportion of the particles residing on the surface will migrate
randomly downward as well, to reenter the exosphere at a later
time. A somewhat similar process has been discussed in \citet{Hodges1982} and
modelled for water ice in polar cold traps by \citet{Schorghofer+Aharonson2014}. 
As is discussed in section~\ref{sec:results:time of day}, 
we find that such a process is required to explain the LADEE
data while maintaining realistic desorption energies.

The persistent reservoir of adsorbed particles 
residing on the surface at night would therefore act as a source for 
building up a distribution of particles with depth by the
end of the night, which then re-emerge during the day. 
Section~\ref{app:model:surface interaction} shows that
these timescales arise naturally from the regolith structure and temperature. 
We assume that particles on the dayside do not adsorb frequently enough 
for long enough to ``squirrel'' in significant numbers.

We use a very simple model to investigate the effect this process has
on the exosphere. Any particle adsorbed to the nighttime surface is
given a small (constant) probability, $P_{\text{sq}}$, of becoming buried 
in the regolith. If this happens, then the particle will reenter the
exosphere at some time during the following day. 
That is, the simulation time for the particle is advanced until the Moon
has rotated it into a random local time on the dayside, with a uniform
probability distribution. The particle ``emerges'' residing on the surface,
then will desorb and hop as normal.

It is likely that this process is much more complicated in reality, with a
dependence on temperature, gradients of temperature and density, and other factors. 
As one such example, just after sunrise, the subsurface is colder than the surface. 
This temperature gradient would discourage particles from migrating upward, 
perhaps delaying the resurfacing of some squirrelled particles until
later in the day than would otherwise be expected. 
Given these uncertain issues, we note the necessary simplicity of this 
model and use it to explore the plausible effects of the process.

The uncertain surface interaction parameters $Q$ and $P_{\text{sq}}$
are varied in the next section, to determine the best values to
describe the LADEE and LACE results.

\section{Results and Discussion}
\label{sec:results}

In this section, we compare our simulations with the
data introduced in section~\ref{sec:data}. First we show how the
treatment of surface interactions affects the variation of argon with
local time of day. Then we investigate the competing hypotheses to
explain the argon bulge over the western maria. In the final part 
of this section, we study the possibility of
seasonal cold traps being the cause of the long-term variation in the
LADEE argon abundance.

\subsection{Distribution with Local Time of Day}
\label{sec:results:time of day}

\begin{figure*}[t!]
	\centering	
	\includegraphics[width=\textwidth]{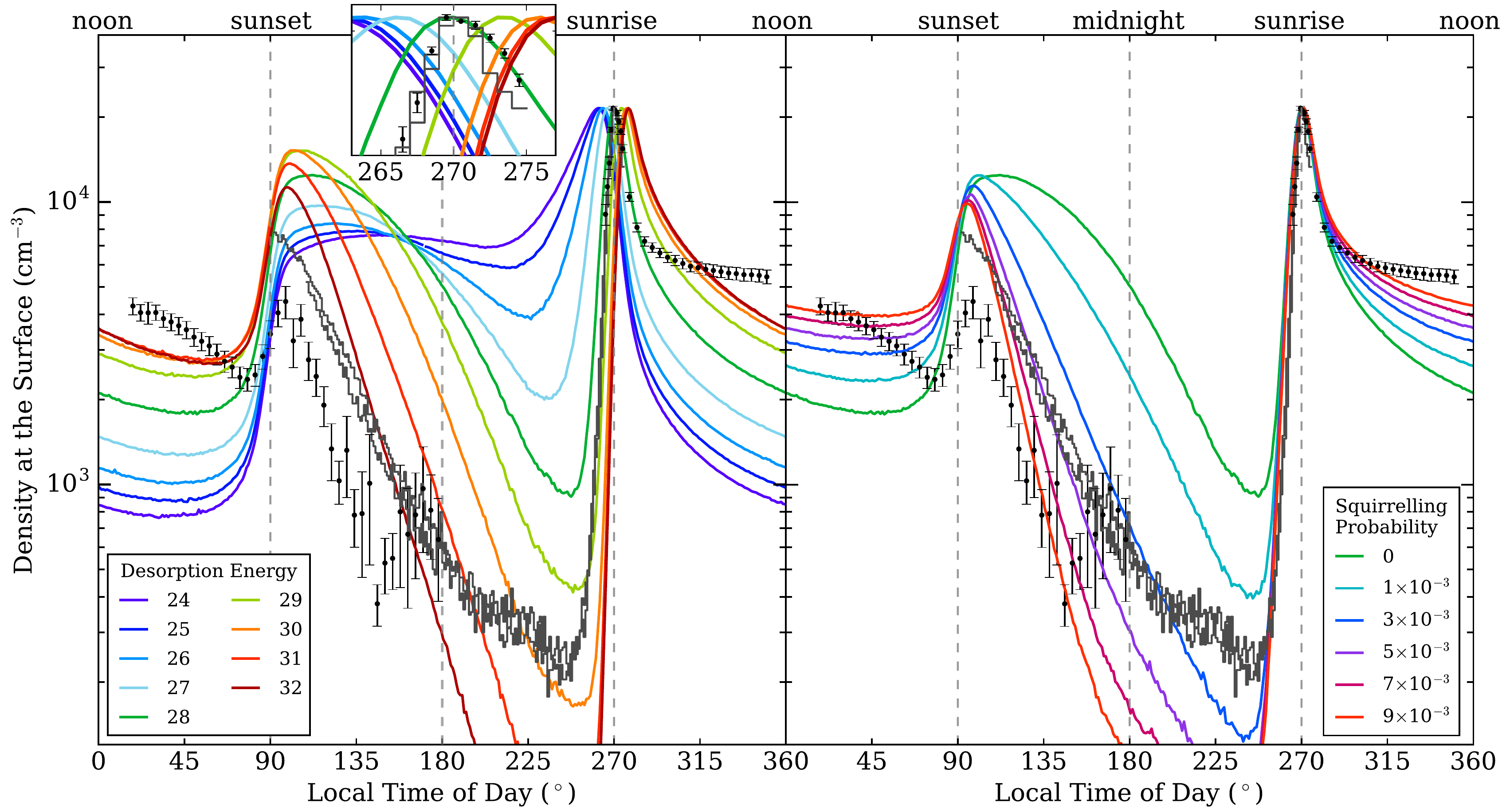}
	\caption{The variation of the argon density as a function of
		local time of day. (left) The desorption energy is varied
		and no squirrelling is included. The labels show the model desorption
		energy, $Q$, in kJ~mol$^{-1}$ and the inset plot zooms in on the region
		around the sunrise peak. (right) ${Q=28}$~kJ~mol$^{-1}$ and the
		model squirrelling probability is varied as shown in the legend. In
		both Figures \ref{fig:local_time_Q_and_P_sq}~(left) and
		\ref{fig:local_time_Q_and_P_sq}~(right), 
		the distributions are normalised to match the peak data
		density to aid comparison of the shape. The mean data and 1$\sigma$ errors from 
		LADEE across all longitudes are shown with black points and 
		the LACE data are shown with grey lines.
		\label{fig:local_time_Q_and_P_sq}}
\end{figure*}

The sensitivity of the simulated exosphere to the values of the model
parameters describing the surface interactions is shown in
Fig.~\ref{fig:local_time_Q_and_P_sq}. In the left panel, the desorption energy,
$Q$, is varied by $\sim$20\% with the squirrelling process switched
off, whereas the right panel shows how the 
distribution changes when the squirrelling probability,
$P_{\text{sq}}$, is altered with fixed ${Q=28}$~kJ~mol$^{-1}$. 

While the higher desorption energy curves
best match the nighttime rate of decrease of argon observed by both
LACE and LADEE, these model surfaces are so sticky that the timing
of the sunrise peak is delayed too far into the day to match the
measurements, as highlighted by the inset panel. The 
sunrise peak position is sensitively dependent upon the model
desorption energy. Given that the sunrise peaks are at the same local
time over the highlands and maria (section~\ref{sec:data:local_time}),
this suggests that the argon interactions with the surface are similar in these different regions. 

Fixing the desorption energy at ${Q=28}$~kJ~mol$^{-1}$ in order to match the
sunrise peak position, the model predicts too much exospheric argon
late in the night and too little during the day. 
This provides empirical motivation for the inclusion of the squirrelling process,
which allows argon atoms to build up a subsurface population in the regolith overnight
that is released throughout the following day. 
Increasing $P_{\text{sq}}$ enables us to produce a
model that matches the nighttime decrease in argon and the shape
of the sunrise peak, as shown in
Fig.~\ref{fig:local_time_Q_and_P_sq}~(right). For our fiducial value of
${P_{\text{sq}}=5\times10^{-3}}$, the simulation agrees within a
factor of 2 with the observations over the entire lunar day. The behaviour
from mid-morning to sunset is somewhat discrepant, but given the
simplicity of our model and the fact that these details do not
change any of the subsequent results and conclusions, we 
do not complicate the model further. 

In contrast to our squirrelling approach, \citet{Hodges+Mahaffy2016}
adopted a bespoke temperature-dependent desorption energy 
(up to $\sim$120~kJ~mol$^{-1}$ at noon) to bring their model into agreement with the LADEE
measurements. Introducing all of this freedom into the model can lead
to a good fit, but, as \citet{Hodges+Mahaffy2016} themselves noted,
such high desorption energies do ``not comport with thermal energies''. 
The energies required by their model during the day are far beyond the bond
strengths that argon has been measured to make or
could be expected to make for the simple van der Waals interactions of a noble 
gas, as discussed in section~\ref{sec:model:surface interaction}. 
Any variable-energy model cannot affect the dayside densities significantly 
without resorting to these extreme values, because the high dayside temperatures
make the residence times negligible for any lower desorption energies.
Thus, the nighttime and sunrise densities cannot be simultaneously matched 
without either including a squirrelling-like process or using unrealistically high,
temperature-dependent desorption energies.
Note that we also tested a similar model for use in the following bulge and long-term 
investigations, and the subsequent conclusions were unchanged.

While our squirrelling approach provides a mechanism for fitting
the nighttime and daytime argon abundances using
physically plausible desorption energies, the simplifications that
this model entails should be noted. These processes have not been extensively studied
with argon on terrestrial materials, let alone in situ \citep{Dohnalek+2002}. 
In reality, there will be a range of
adsorption sites with somewhat different desorption energies, and the
probability of adsorption will vary depending upon both the speed of
the incoming atom and the presence of volatiles already on the surface.
For instance, experiments 
show that argon has about a 70\% probability of adsorbing to a
surface at typical exospheric impact speeds of
300~ms$^{-1}$ but much lower temperatures of
22~K \citep{Dohnalek+2002}. This probability decreases rapidly with
higher impact speeds, reaching zero for 550~ms$^{-1}$. Argon is
also more likely to stick when other argon atoms are already residing on the
surface \citep{HeadGordon+1991}. 
The 300~ms$^{-1}$ adsorption probability reaches one when the
coverage is roughly a monolayer. These values may of course be
somewhat different for adsorption to the lunar regolith. 
Note also that the value of $Q$ depends on the exact form of Eqn.~\ref{eqn:t_res}. 
So, it should not be interpreted as a precise estimate of the true energy, 
especially given the aforementioned details that are all approximated 
into this single parameter. 
For example, \citet{Grava+2015} used a different prefactor and an extra free parameter, 
so their fitted value of $Q=27$~kJ~mol$^{-1}$ results in a curve with the sunrise 
peak around 275$^\circ$, comparable to our $Q=30$~kJ~mol$^{-1}$. 
We focussed on fitting the observed sunrise peak time at 270$^\circ$, 
so find an effectively lower energy.

We ran additional simulations to investigate the sensitivity of our 
results to these adsorption issues. Lowering the adsorption probability
has a similar effect to lowering the desorption energy. Even for an 
adsorption probability below 0.1, an increase of a few kJ~mol$^{-1}$ to $Q$ results in 
a similar distribution and sunrise peak position. A speed-dependent adsorption 
probability also does not dramatically change the distribution, compared with the 
effects from changing $Q$ or $P_{\text{sq}}$. 
More importantly, no such mechanisms appear to reduce the need for the 
squirrelling process to match the high dayside densities. So, while these known 
simplifications affect the results at a level that could explain
some of the small discrepancies between the model and data distributions,
our main conclusions are not sensitive to these assumptions. 

One final consideration is the cold trap area and corresponding source rate. 
The above simulations were all run with the low permanent cold trap fractional 
area of 0.01\% (of the polar 15$^\circ$) that approximately corresponds to the 
low source rates estimated by \citet{Killen2002}.
The much larger traps and high rates considered in the next
section have the effect of increasing the density toward the end of the night, which 
improves the match to the minimum LACE densities 
(although these may be below LACE's sensitivity \citep{Hoffman+1973}), 
due to the emergence of newly created particles through the night. 
The distribution is unaffected at other local times of day.

\subsection{Longitudinal Bulge}
\label{sec:results:bulge}

There are two alternative hypotheses for the origin of the argon bulge
over the western maria: (1) it reflects a spatially variable source rate
that is higher over the western maria \citep{Benna+2015}; 
or (2) it reflects a lower desorption energy
from the western maria \citep{Hodges+Mahaffy2016}. We perform
simulations using either a localised source reflecting the LPGRS potassium
map (and our ${Q=28}$~kJ~mol$^{-1}$, ${P_{\text{sq}}=5\times10^{-3}}$ 
model described in section~\ref{sec:results:time of day}), 
or a uniform global source and a spatially varying desorption energy 
to examine these two scenarios. 

For the case of a local source with a greater release of argon over
the western maria, the amplitude of the bulge depends
sensitively on (1) how localised the source is and (2) the source rate, 
or -- equivalently in the steady state -- the loss rate. 
For a diffuse source, a rapid turnover of particles with short lifetimes 
is necessary to produce argon atoms
that have not travelled so many times around the Moon that their
locations no longer reflect their origin. We ran simulations with a
local source (proportional to the LPGRS potassium abundance) 
with a range of cold trap fractional areas to determine the area
required to produce a bulge comparable with that observed by LADEE. 

\begin{figure}[t!]
	\centering	
	\includegraphics[width=\columnwidth]{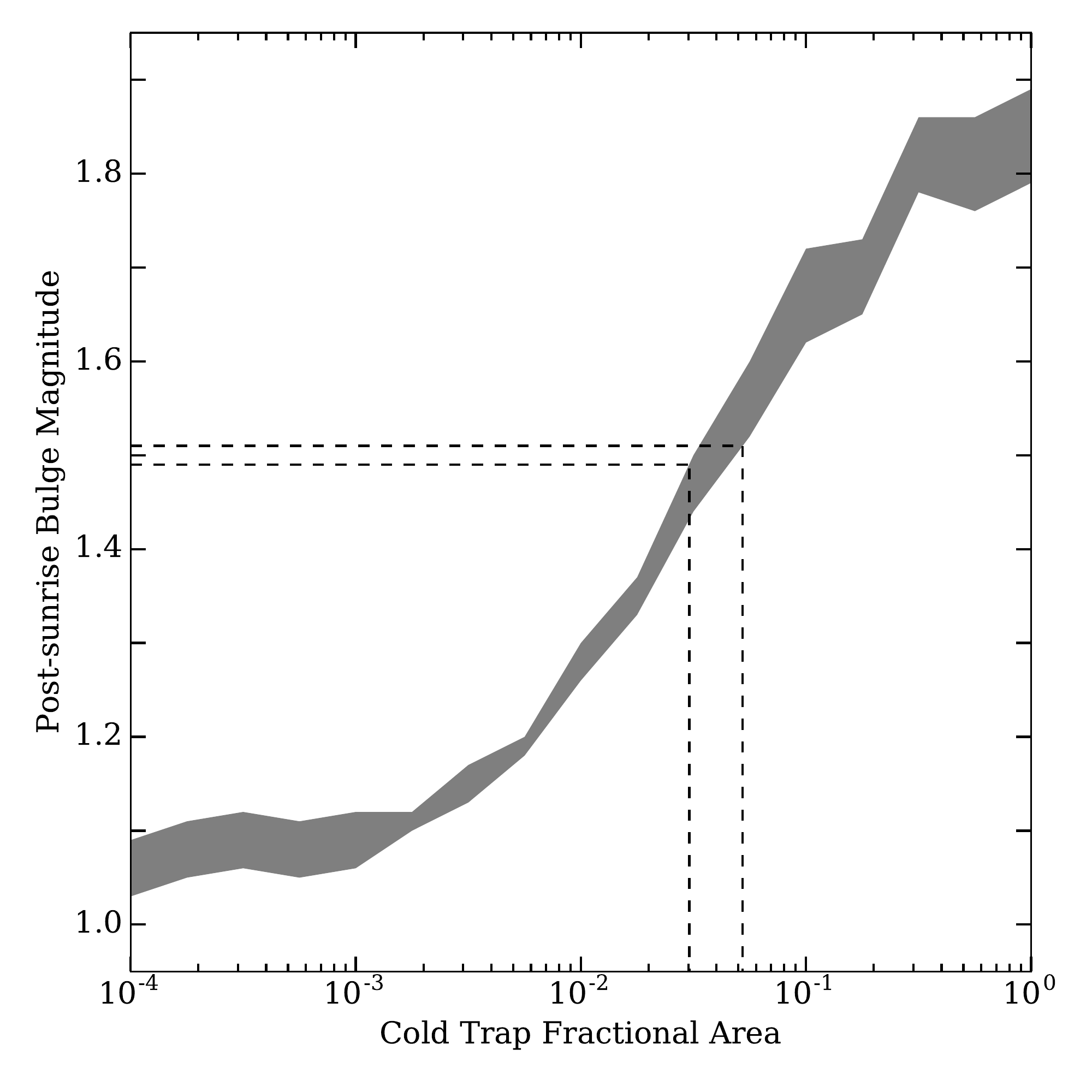}
	\caption{The magnitude (the ratio of the maximum to minimum densities 
		over all longitudes at a given time of day) of the 
		simulated bulge over the western maria
		(in this case at 280$^\circ$--$300^\circ$ local time of day), generated by the
		localised source, for different values of the cold trap fractional area of
		the polar 15$^\circ$. The dashed lines show the magnitude and uncertainty of the 
		bulge measured by LADEE at the same local time and the corresponding 
		range of cold trap areas.
		\label{fig:bulge_magnitude_from_cold_traps}}
\end{figure}

Fig.~\ref{fig:bulge_magnitude_from_cold_traps} shows the amplitude of the
post-sunrise (280$^\circ$--$300^\circ$) argon bulge over the western
maria. As shown by the dashed lines, to match the LADEE maximum-to-minimum ratio of ${1.5\pm0.1}$, 
a cold trap fractional area of ${4\pm1}$\% of each polar 15$^\circ$ is necessary, 
corresponding to a mean lifetime of $\sim$1.4~lunar~days. 
This area is comparable
with the $2$--$2.5$\% covered by PSRs \citep{Mazarico+2011}, 
but is uncomfortably large if argon is
only supposed to be permanently trapped in regions with temperatures
never exceeding $\sim$40~K \citep{Hodges1980b}.
As indicated in Fig.~\ref{fig:lifetime_from_cold_traps}, 
cold traps of this large size appear to correspond to regions with 
temperatures that can reach as high as 175~K. 

Therefore, for this to be a viable model, 
one would either need argon to be more readily lost from
the exospheric system than previously anticipated, 
or to have a more highly localised source below the surface 
than the LPGRS near-surface potassium map. 
To further investigate the degeneracy between the source rate
and how localised the source is, we also tested a ``top-hat'' and a point source
in the same way. The top-hat source emits argon uniformly from all regions 
with at least 2,000~ppm of potassium, giving a source region 
covering 6\% of the Moon's surface area in the PKT. 
This can create an argon bulge with the required 
amplitude with a cold trap fractional area of only 0.4\%, 
a lifetime of 5.3~lunar~days, and a source and loss rate of 
${2.9\times10^{21}}$~atoms~s$^{-1}$. 
For the extreme case of a point source at 335$^\circ$ longitude on the equator, 
a cold trap fractional area of 0.2\% is sufficient, with a lifetime of
8.1~lunar~days and a rate of ${1.9\times10^{21}}$~atoms~s$^{-1}$.

If the source is not quite so localised, then feasible causes of increased losses might be: 
an abundance of small-scale cold traps such as those inferred
on Mercury \citep{Paige+2014,McGovern+2013}; 
the presence of other volatiles in PSRs
increasing the adsorption probability \citep{Dohnalek+2002}; 
and an unaccounted-for loss mechanism that means that our assumed solar radiation 
loss rate is an underestimate 
(recall the large uncertainty on this value as mentioned in section~\ref{sec:model:losses}). 
Thus, this model begs an explanation either for the high
source and loss rates, or for a highly localised source. Noting this tension, we
press on to investigate the shape of the argon bulge in the simulation from
the localised (potassium-weighted) source with a cold trap fractional polar area
of 4\% and compare it with the data.

\begin{figure}[t!]
	\centering	
	\includegraphics[width=\columnwidth]{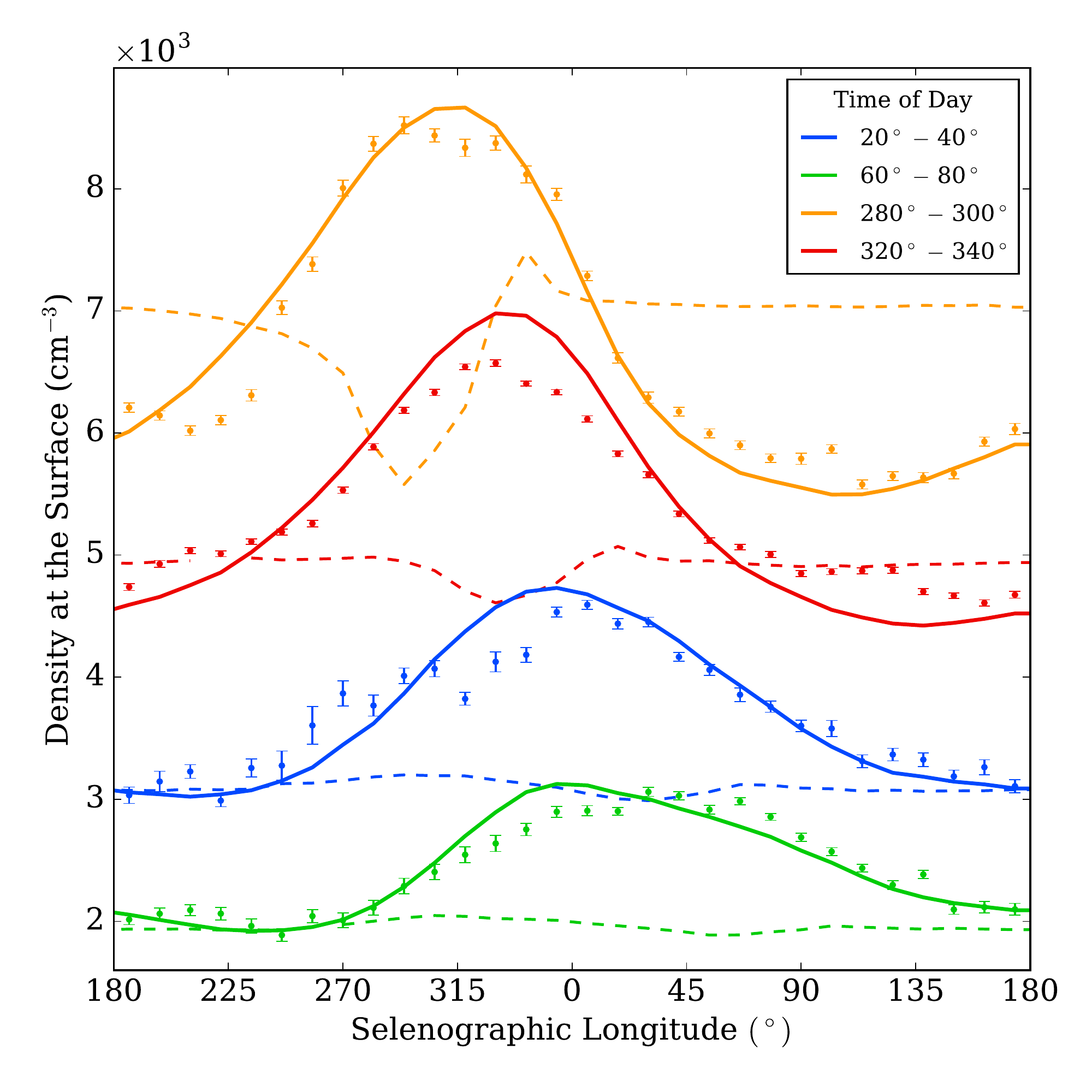}
	\caption{The variation of argon density with selenographic longitude for a 
		representative selection of local times of day, shown by the
		different colours as defined in the legend.
		The solid lines show the bulge from the local source model with 
		high rates of source and loss, 
		the LADEE data are shown as points, 
		and the dashed lines show the results for the global 
		source with a lower desorption energy in the mare region than in the 
		highlands (26 and 28~kJ~mol$^{-1}$ respectively).
		\label{fig:bulge_results}}
\end{figure}

The variation of argon density with longitude in our simulations is shown in
Fig.~\ref{fig:bulge_results} for a sample of different local times of day. 
The solid lines result from the local source and large cold traps. 
The dashed lines are from the alternative hypothesis of an isotropic 
source with mare and highland desorption energies of 
${Q=26}$~kJ~mol$^{-1}$ and 28~kJ~mol$^{-1}$ respectively, with a
low cold trap fractional area of 0.01\% (of the polar 15$^\circ$) 
that corresponds approximately to the
source rate estimated by \citet{Killen2002}. (Note that with an isotropic source
the shape of the bulge is insensitive to the source rate.) Also reproduced are
the LADEE results. We are predominantly interested in the shapes of the
different curves, and not their relative amplitudes, which are determined by the 
local time of day distribution and are slightly different at certain times, 
as discussed in section~\ref{sec:results:time of day}. 
Thus, for clarity, each model curve was divided by its mean and 
multiplied by the mean of the corresponding data set.

The model with the local source was tuned only to match the
maximum-to-minimum ratio of the post-sunrise argon bulge over the western
maria observed by LADEE. However, the relative sizes of the bulge at all other times of day,
the position, width, and shape of the bulge, and the shift of the bulge to the east 
throughout the day also happen to be reproduced well. These
features result from the fact that the overnight 
build-up of argon over the western maria around longitudes 
$\sim$300$^\circ$--$330^\circ$ diffuses rapidly across the
sunlit surface after it reaches sunrise. This diffusion spreads out
the peak in the argon density and shifts it into the dayside: that is,
to the east. All these features are also reproduced with the top-hat 
and point sources, apart from the point source bulge being slightly narrower.

In contrast to the successes of this local source model, the effect of
decreasing the adsorption energy over the western maria,
shown by the dashed lines in Fig.~\ref{fig:bulge_results}, does
not match any of the features in the data. The failure to reproduce the
observations arises because the lower desorption energy 
encourages atoms to leave the surface and hop more frequently. 
This is successful in producing a bulge towards the end of the night,
where the residence time is longest.
However, it also rapidly evacuates the argon from the maria, and the nighttime
bulge is replaced by a deficit in argon density almost immediately after sunrise.
Therefore, trying to create a local overdensity in this way will inevitably fail if the
bulge is required to persist throughout the day, as it is observed to do. 

A separate reason to doubt this explanation is that these small changes
in adsorption energy lead to significantly different local times for
the sunrise peak in the mare and highland regions, in contrast to what the
LADEE data show. Consequently, a spatially varying desorption energy
explanation for the argon bulge can be ruled out for a couple 
of reasons. Similar arguments can be used to dismiss the idea that the
bulge results from hotter surface temperatures for the maria, for example. 

The local source is thus the only proposed hypothesis that has the potential to
reproduce the variation of argon density with selenographic longitude
seen in the data, and the results are remarkably similar. However, for this 
explanation to be successful, either (1) the lifetime of an argon atom in the
exospheric system (i.e. from source to sink) needs to be $\sim$1.4~lunar~days -- 
if the source rate is proportional to the LPGRS potassium
abundance; or (2) the source must be highly localised 
(or a slightly less extreme combination, as illustrated by the top-hat model). 
For the diffuse source, the required source rate of 
${1.1\times10^{22}}$~atoms~s$^{-1}$ is 
about 46\% of the total rate of argon produced in the Moon \citep{Hodges1975},
which is unlikely to be able to reach the surface unimpeded.
The correspondingly high loss rate that this implies appears to demand widespread polar
cold-trapping of argon that exceeds what had previously been considered.
Assuming that the cold traps have been stable for $\sim$1~Gyr \citep{Arnold1979}, 
this means that the order of $10^{13}$~kg of argon 
would have been delivered to the polar regions during this time.

With this high cold trap fractional area of 4\%, 
it takes roughly six~lunar~days from the start of the simulation 
before the exosphere reaches an equilibrium of source and loss rates
and a stable number of argon atoms. 
In comparison, the time it takes for the equilibrium shape of the distributions 
to become established is always very short, on the order of one~lunar~day. 
This timescale is the same even with much smaller cold traps 
(such as those required for a highly localised source), 
for which the system can take a long time to reach a true steady state. 
This is analogous to saying that a localised event diffuses rapidly across the system, 
even if the total number of particles is still offset from equilibrium.

Irrespective of the lifetime and loss rate of argon in the simulation,
particles spend 60\% of their life residing on the surface, 
30\% squirrelling under the surface, 
and only 10\% in flight in the exosphere.
Thus, at any given time, these same proportions of the 
population of argon atoms will be residing on the surface, 
squirrelling, and in flight.
The total number of argon atoms in the simulated exosphere at any time is about
${4\times10^{28}}$, corresponding to a mass of 2,600~kg.

\subsection{Long-Term Variation}
\label{sec:results:long-term}

\citet{Hodges+Mahaffy2016}
suggested that seasonally varying cold traps could produce a periodic
signal responsible for the smooth long-term variation
in the argon exospheric density by 28\% over the 5 months of LADEE's
operation. As described in section~\ref{sec:model:seasonal}, 
we have included seasonal cold traps
to account for the fact that due to the 1.54$^\circ$ obliquity of the Moon,
when one pole is tilted away from the Sun, larger areas will act as
cold traps for a few months. These seasonal traps would both
temporarily remove and add argon, so they could drive
changes in the density on the relevant timescale without a change in
the overall source and loss rates. 
We performed simulations to test this hypothesis. 

The magnitude of this variation is affected by the peak size of the seasonal traps
and also by the source and loss rates, which in our model are controlled by the sizes
of the permanent cold traps. For the low permanent cold trap fractional 
area of 0.01\% (of the polar 15$^\circ$) that corresponds approximately 
to the low source rate estimated by \citet{Killen2002} 
(where solar wind losses dominate), 
the 28\% change in the density seen by LADEE was reproduced
with a peak seasonal cold trap fractional area of 
${f_{\text{peak}}=0.8}$\% at each pole's midwinter.

\begin{figure}[t!]
	\centering
	\includegraphics[width=\columnwidth]{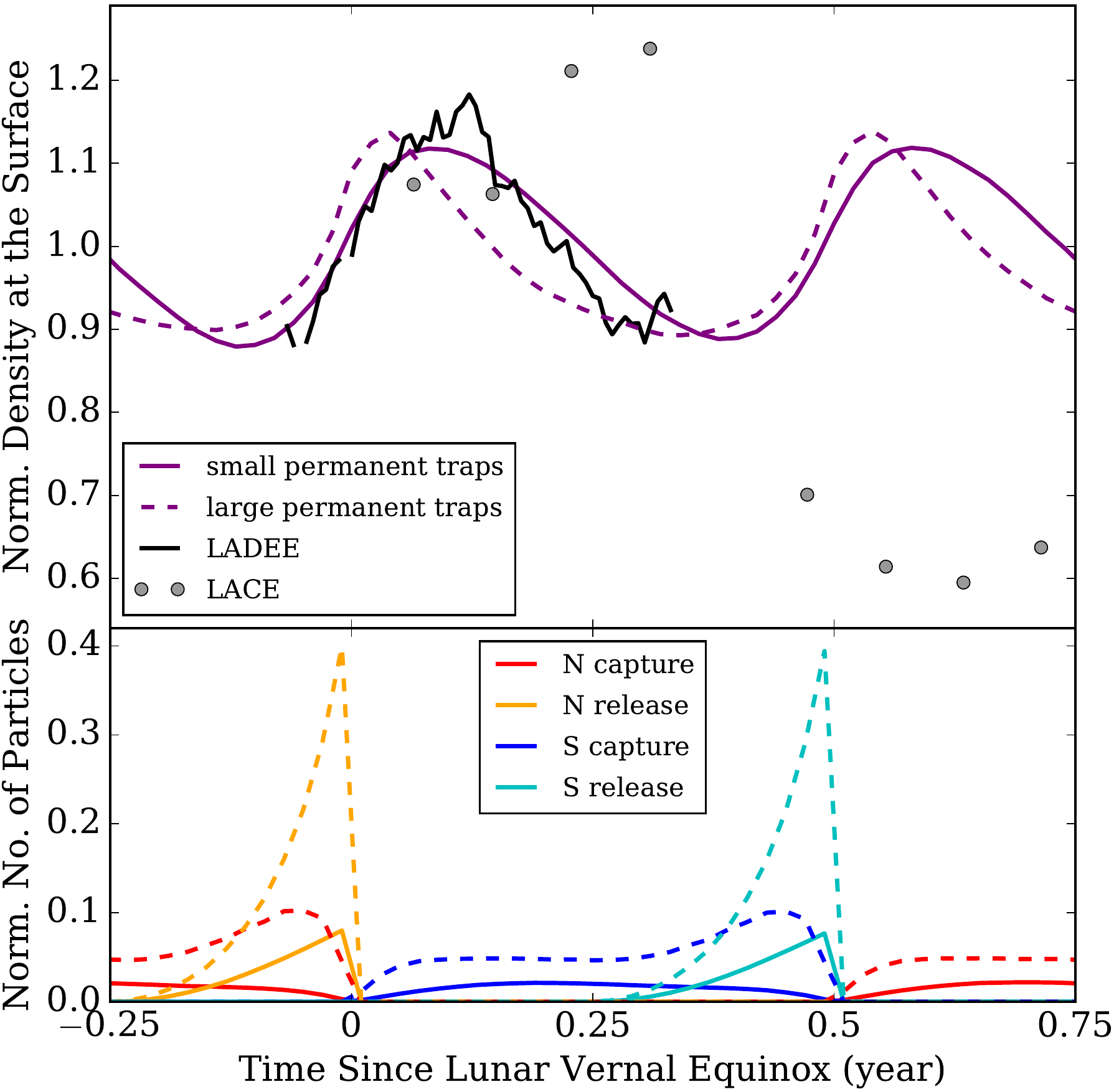}
	\caption{(top) The long-term variation of the argon population
		within 30$^\circ$ latitude of the equator, normalised by
		the mean density. The purple solid line shows the normalised 
		simulated density at the surface for
		the low rates model with permanent and peak seasonal cold trap fractional areas 
		of 0.01\% and 0.8\% respectively; and the dashed line for a high rates simulation 
		with areas of 4\% (which reproduced the bulge for the localised potassium-weighted 
		source) and 16\%.
		Time is measured from the last lunar vernal equinox.
		The LADEE data are shown by the black line for comparison, 
		and the grey points show the magnitudes of the LACE sunrise peak  densities 
		\citep{Hodges1975} relative to the mean sunrise density measured by LADEE.
		(bottom) The number of argon particles that are 
		being captured and released in each simulation by the 
		northern and southern seasonal traps, 
		given as a fraction of the total number of argon atoms in the 
		equilibrium system (${4\times10^{28}}$). The dashed lines show the results for the large permanent traps. 		
		\label{fig:long_term_results}}
\end{figure}

The periodic long-term variation that is produced by these
seasonal traps, at the latitudes probed
by LADEE, is shown by the solid lines in Fig.~\ref{fig:long_term_results}, over a
period of 1~year in the simulation. The peak density is predicted to be delayed by about
0.07~years (1~lunar~day, $\sim$1~month) after the minimum trap
size at the equinox. This is related to the time it takes for argon to travel between 
the poles and the equator and is remarkably similar to the delay in the 
LADEE data after the lunar vernal equinox, as shown in Fig.~\ref{fig:long_term_results}.
Also shown is how much argon becomes
trapped and released at each point in time throughout the
year by the seasonal traps. Particles can be trapped from the beginning of winter 
until the very end, but are only released starting after midwinter,
when the traps begin to shrink. This leads to a mild asymmetry in the
long-term variation, which could easily be modified by deviating from our simple 
sinusoidal model.

To demonstrate how the efficacy of the seasonal traps depends on the 
argon lifetime, we can attempt to reproduce the long-term variation 
with the very large permanent trap areas (4\%) and higher loss rates that, 
with a localised source weighted by the near-surface potassium, 
would produce a bulge similar to the LADEE data. 
We found that much larger seasonal traps would be needed to cause
the same magnitude long-term variation, 
with a peak fractional area of $f_{\text{peak}}=16$\%. 
This is because, with much shorter particle lifetimes, 
fewer particles near the equator have been affected by the seasonal traps. 
In this case, a maximum of about ${2.6\times10^{28}}$ 
argon atoms (over half of the steady-state exosphere) 
must be temporarily trapped to effect the 28\% variation 
near the equator, over twice as much as in the low loss case. As shown by 
the dashed line in Fig.~\ref{fig:long_term_results}~(top), the delay of the peak density
is also affected and occurs 0.04~years (one half~lunar~day) earlier. 
This timescale is sensitive to the evolution of the seasonal traps.
It is possible that unaccounted-for factors such as thermal inertia, which could cause 
newly shadowed regions not to begin trapping immediately, delay
the variation enough to reproduce the data even with high loss rates.

Unfortunately, the success of this seasonal model with the LADEE data does not
continue for the long-term variation measured by LACE, which is
shown by the grey points in Fig.~\ref{fig:long_term_results} (assuming it is 
not the result of instrument degradation). 
The seasonal variation acts in the opposite way to the trend seen in 1973. 
Therefore, if there are seasonal cold traps that 
explain the LADEE argon data and they were active during the LACE measurements, 
then the loss rate required to
match the drop measured by LACE needs to be significantly larger than had been
anticipated by \citet{Grava+2015}. It is also noteworthy that the later
LACE measurements fall well below the minimum measured by LADEE at the same location.
This suggests that, if the LADEE data do show a periodic feature and there was not 
a transient loss event to drive down the later LACE measurements, then the equilibrium 
size of the exosphere must have increased over the last 40 years. 

It is also possible that the LADEE and LACE long-term variations are both the result of 
transient source events, as suggested by \citet{Benna+2015} and \citet{Grava+2015}. 
If we assume that the minima of the LADEE and LACE measurements indicate the equilibrium 
states throughout the measurement periods, and that a transient source
had increased the density to the maximum before discontinuing, then we can model the
decrease as a simple exponential decay from the maximum measurement 
back to the equilibrium state.
In both the LADEE and LACE cases this would require a 
lifetime of around 0.9~lunar~days, even shorter than that required for our
local source bulge hypothesis and implying even greater loss rates.
If the equilibrium level is lower than the minimum measurements, then the 
variations would therefore be part of even larger but less rapid declines.
In the absolute limit of no background exosphere at all, the required lifetimes 
could extend up to nine and five~lunar~days for LADEE and LACE respectively. 
In this extreme case, these long lifetimes would require only small source and loss rates 
and correspondingly small permanent cold traps with temperatures below 70~K, 
as shown by Fig.~\ref{fig:lifetime_from_cold_traps}.
Similar arguments can be made
regarding transient loss events, since LADEE shows an equally rapid increase of argon. 
Of course, it is imaginable that a combination of multiple, dramatic source and 
loss events could produce these variations regardless of the lifetime, 
but this is extraordinarily unlikely.

As discussed in section~\ref{sec:model:seasonal}, 
asymmetrical variation of the seasonal traps in summer and winter 
is necessary to match the half-year period suggested by the data. 
Therefore, our model included no seasonal cold traps throughout the 
summer half of the year at each pole. Had we instead allowed some seasonal traps 
to decrease until the summer solstice, 
their reduction would have offset some of the effect of the 
growing traps at the other pole. 
Consequently, the model would have needed larger seasonal trap areas 
to match the amplitude of the observed long-term argon variation.
It should be possible to use lunar elevation maps to predict the 
actual variation of seasonal cold traps throughout the year,
at least enough to test whether such an asymmetrical variation 
is realistic. In the scope of this work, we show only that
this hypothesis has the potential to explain the data.

\section{Conclusions}	
\label{sec:conclusions}

We have studied the LADEE (and LACE) measurements of the lunar argon exosphere
and developed a Monte Carlo model to investigate what is implied about
the sources, sinks, and surface interactions in the system, 
and to test whether various hypotheses are able to explain the observed features. 

The extrapolation of simulated density at an altitude of 60~km to density
at the surface is fitted to within 12\% everywhere using a
model of two Chamberlain distributions with different
temperatures. From this altitude, using a single Chamberlain distribution or its first
order approximation can lead to overestimates greater than a factor of
3. These errors can be much larger for extrapolations from higher
altitudes. 
Other exospheric species should exhibit similar behaviour. 
Lighter particles typically travel farther each hop, which would 
increase the error from using a single-temperature model.
The two-temperature model fits the LADEE data well, suggesting 
that simple thermal desorption dominates the release energetics of exospheric argon.

The distribution with local time of day of the argon density in the exosphere
is very sensitive to the nature of the interactions with the surface. 
Apart from an offset in amplitude reflecting the higher density over
the maria, the highland and mare results are very similar, suggesting
that the surface interactions do not differ greatly with regolith
composition at equatorial latitudes. 
A very simple model allowing atoms to squirrel into the regolith
overnight, building up a subsurface population that is released during the
following day, can reproduce the broad
characteristics of the observed exosphere at all times of day, 
without the need to resort to unreasonably high and 
temperature-dependent desorption energies.
The timing of the sunrise peak requires a residence time near
sunrise of 1,300~s, which corresponds to a desorption energy of
28~kJ~mol$^{-1}$, a high but plausible value for noble gas interactions.
The subsequent results are insensitive to the
details of these surface interaction models.

Of the two hypotheses that have been proposed for the origin of the
argon bulge over the western maria and PKT, only a localised source
has the potential to explain this feature. 
Our simulations with this model can reproduce the observed size, shape, and 
position of the bulge at all local times of day. There
is a degeneracy between how localised to the mare region the source is
and the lifetime and rates that the data require. For a source distribution
weighted by the LPGRS potassium map, the observed bulge is reproduced
with a mean lifetime for argon atoms in the exospheric system of only
1.4~lunar~days, corresponding to a high equilibrium source and loss rate
of ${1.1\times10^{22}}$~atoms~s$^{-1}$. 
To achieve this, our model would need permanent cold
traps that have a total area comparable with the PSRs measured at Diviner's
resolution, or some other additional loss mechanism. 
A more highly localised source can reduce the required rates and
trap areas by an order of magnitude -- a point source reproduces a bulge of the right
amplitude with a source and loss rate of ${1.9\times10^{21}}$~atoms~s$^{-1}$.
So, despite this model's unique success in reproducing the data, 
it begs an explanation for some combination of source localisation
and high source and loss rates.

Models that aim to create the argon bulge by
encouraging atoms to hop either more frequently or higher founder
because they naturally lead to a short-lived feature through the night that
is replaced by a local deficit in the argon density after sunrise.

The long-term variation in the global argon density seen by LADEE can
be elegantly explained by the periodic behaviour of seasonal cold
traps. The details of how large they need to be depend upon the 
base source and loss rates. The time lag of the peak
density in the data is reproduced naturally by the model for small cold traps
and low rates. It is slightly offset for higher rates, which might be 
mitigated by effects such as thermal inertia. 
However, the long-term decrease seen by LACE in 1973,
if real, requires some other significant source and/or loss process 
because the seasonal variation should
act in the opposite way to the observed trend.
The relatively smooth variation of the argon density observed over the lifetime of
LADEE suggests that significant transient release or loss events are unlikely
to be the cause. 
This includes the apparent lack of a significant effect from the periodic crossing of the Moon through the Earth's magnetotail, which might have been expected to reduce the solar wind loss rate.
Any transient source (or loss) explanation would also require high rates of source and loss for the 
system to return to equilibrium after the event within the measured lunar-day timescales,
unless the equilibrium density is far lower than the minimum observed by LADEE.

Seasonal cold traps should be expected to impact other species in
the exosphere in a similar way, depending on their threshold trapping
temperature. If any non-radiogenic, condensible species 
(such as methane) \citep{Hodges2016} were also found to
follow the variation seen for argon, then this would be strong
evidence in support of the seasonal hypothesis (and vice versa). 
This is because tidal or seismic changes
that might affect the argon source rate would be irrelevant for species that
do not come from inside the Moon. Further long-term observations of the 
argon density would also help determine whether the variation is actually periodic
in the first place.

There are several experiments that could help determine
what combination of source localisation and rate of source and loss
is responsible for the bulge, given the lack of other possible explanations.
To test the hypothesis of a diffuse localised source 
with very high source and loss rates, one could pursue: 
in-situ searches for argon trapped in PSRs, 
although there are various uncertainties regarding how much sequestered argon would 
be found and at what depth \citep{Schorghofer+Taylor2007}; 
or measurements of the exosphere towards the poles,
where the effects that large cold traps would have on the distribution of argon 
with latitude would be detectable. 
On the other end of the degeneracy: if the source is highly localised, then
the large differences in that rate should directly affect the late-night argon 
distributions in the mare and highland regions. This might need to be measured at
or near the surface to detect the very low densities. 
Future investigations of this kind would help determine if 
this model is indeed the origin of the bulge, or if
some entirely new explanation is required.


%
%
%
\appendix

\section{Theory and Derivations}
\label{app:model}
This appendix contains the derivation of the hop trajectory and
time of flight equations. Also included are details 
of the ``simulation-fit'' altitude model described in section~\ref{sec:data:altitude} 
and a short calculation concerning the
squirrelling mechanism described in section~\ref{sec:model}.

\subsection{Notation}
\label{app:model:notation}
\begin{figure*}[t!]			
	\centering
	\begin{subfigure}[t]{0.59\textwidth}
		\includegraphics[width=\textwidth]{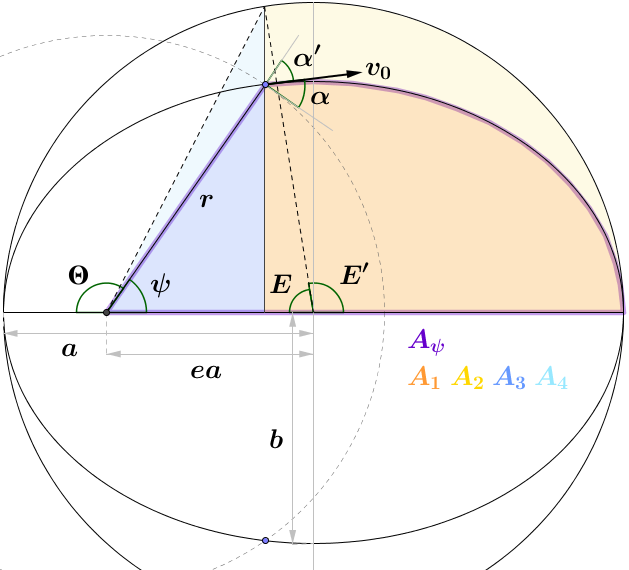}
		\subcaption{Ellipse and circumcircle notation.
			\label{fig:ellipse_derivations}}
	\end{subfigure}		
	\begin{subfigure}[t]{0.39\textwidth}
		\includegraphics[width=\textwidth]{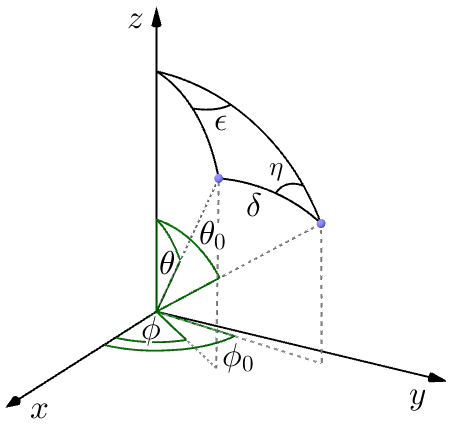}
		\subcaption{Start and end point notation.
			\label{fig:particle_bounce_notation}}
	\end{subfigure}
	\caption{The notation for the dimensions, angles and areas:
		\textbf{(a)} $a, b,$ and $e$ are the semimajor axis, semiminor
		axis, and eccentricity of the ellipse, respectively. The variables 
		$\theta$ and $E$ are the true
		anomaly and mean anomaly of the object at a distance $r$ from the
		focus; $\psi$ and $E'$ are their complementary angles. The $\alpha$ and
		$v_0$ give the velocity of the object; $\alpha'$ is the
		complementary angle. $A_{\psi}$ is the area of the outlined triangle
		$+$ ellipse section; $A_1, A_2, A_3$, and $A_4$ are the areas of the
		same-colour-shaded sections. The Moon is shown by the dashed circle;
		the blue points mark the start and end of a particle's
		hop. \textbf{(b)} The subscript 0 denotes the initial
		position. The $\eta$ is the direction west from north of the particle's
		initial velocity. The $\epsilon$ is the difference in longitude and
		$\delta$ is the total angle between the start and end points. 
		\label{fig:notation}}
\end{figure*}

Figs.~\ref{fig:notation}a and b show all the relevant notation 
for a particle's hop. We use standard
polar coordinates $\theta$ and $\phi$, with colatitude $\theta=0$ at
the north pole, and standard selenographic longitude $\phi=0$ at
the sub-Earth point. The local time of day,
$\phi'$, is given by the longitude relative to the subsolar point
(noon, where $\phi'=0$). To account for libration, the longitude of
the subsolar point, $\phi_{\text{ss}}$, is calculated from JPL HORIZONS 
ephemerides \citep{Giorgini2015} (latitude variations are ignored). The time
of day is then simply \({\phi'=\phi-\phi_{\text{ss}}}\). If we had
ignored libration, then errors of over 3$^\circ$ would have been
introduced for some local times of day. 

\subsection{Hop Trajectory}	
\label{app:trajectory}

Starting from the initial position and velocity of a particle, we can
calculate the landing position (or the position at any altitude) by
first finding the parameters of the elliptical path, shown in
Fig.~\ref{fig:notation}a. The vis viva equation gives the
semimajor axis, $a$, in terms of the speed, $v$: 
\begin{linenomath*}
	\begin{equation}
	v^2 = GM\left(\dfrac{2}{r}-\dfrac{1}{a}\right)
	\label{eqn:vis viva} \hspace{1em} \Rar \hspace{1em} a = \left(\dfrac{2}{r}-\dfrac{v^2}{GM}\right)^{-1} \;,
	\end{equation}		
\end{linenomath*}	
and the ellipticity, $e$, is found from the velocity angle $\alpha'$:
\begin{linenomath*}
	\begin{align}
	\sin(\alpha') &= \dfrac{r \dot{\psi}}{v} = \sqrt{\dfrac{a^2(1-e^2)}{2ar-r^2}} \\
	\Rar \hspace{2em} e &= \sqrt{1-\left(\dfrac{2ar-r^2}{a^2}\right)\sin^2(\alpha')} \;\;.			
	\label{eqn:e=}
	\end{align}
\end{linenomath*}
The equation for an elliptical path can then be rearranged to give the
angle $\psi$ in terms of $a$ and $e$: 
\begin{linenomath*}
	\begin{align} 
	r(\psi) &= \dfrac{a(1-e^2)}{1-e\,\cos(\psi)}
	\label{eqn:elliptical path}\\
	\Rar \hspace{2em} \psi &= \cos^{-1}\left[\dfrac{1}{e}\left(1 - \dfrac{a(1-e^2)}{r}\right)\right] \;. 
	\label{eqn:psi=}
	\end{align}	
\end{linenomath*}
In Fig.~\ref{fig:notation}b the angles $\delta$ and
$\epsilon$ can be calculated along with the landing coordinates
$\theta$ and $\phi$. At the start of the hop (${\psi = \half
	\delta}$), Eqn.~(\ref{eqn:psi=}) becomes 
\begin{linenomath*}
	\begin{equation}
	\delta =  2\cos^{-1}\left[\dfrac{1}{e}\left(1 - \dfrac{a(1-e^2)}{R}\right) \right] \;,
	\end{equation}
\end{linenomath*}
where $R$ is the radius of the Moon.

The spherical cosine rule gives
\begin{linenomath*}
	\begin{equation}
	\begin{split}
	\theta=\cos^{-1} & \left[ \cos(\theta_0) \cos(\delta)\; + \right.  \\
	& \,\left. \sin(\theta_0) \sin(\delta) \cos(\eta) \right] \;. \label{eqn:end theta}
	\end{split}
	\end{equation}
\end{linenomath*}
Using the cosine rule again, but with $\epsilon$ instead of $\eta$, gives 
\begin{linenomath*}
	\begin{equation}
	\epsilon = \cos^{-1} \left(\dfrac{\cos(\delta) - \cos(\theta)\cos(\theta_0)}{\sin(\theta) \sin(\theta_0)} \right)\;. \label{eqn:epsilon}
	\end{equation}
\end{linenomath*}
Then $\phi$ is simply given by 
\begin{linenomath*}
	\begin{equation}
	\phi = \begin{cases} 
	\phi_0 + \epsilon \hspace{2em} \eta > \pi \\
	\phi_0 - \epsilon \hspace{2em} \eta < \pi 
	\end{cases} \!\!\!. \label{eqn:end phi}
	\end{equation}
\end{linenomath*}
The total time of flight is ${t = 2 t_\psi}$, where $t_\psi$ is
the time the particle spends tracing out the area $A_\psi$ in
Fig.~\ref{fig:notation}a with its radial vector. Using general
properties of ellipses, it can be shown that 
${A_\psi \equiv A_1+A_3 = \tfrac{b}{a}(A_2+A_4)}$, and therefore 
\begin{linenomath*}
	\begin{equation}
	A_\psi = \half \,ab\,\left(E'+e\,\sin(E')\right)
	\label{eqn:area} \;.
	\end{equation}
\end{linenomath*}
The total period of the orbit, $T$, is given by Kepler's third law:
\begin{linenomath*}
	\begin{equation}
	T=\sqrt{\dfrac{4\pi^2a^3}{GM}} \;. 
	\label{eqn:Kepler 3}
	\end{equation}
\end{linenomath*}
Kepler's second law states that equal areas are swept out by the
radial vector in equal times, therefore,
\begin{linenomath*}
	\begin{equation}
	\begin{split}
	t_\psi &= T\dfrac{A_\psi}{A_{\text{ellipse}}} = T\dfrac{\half ab(E'+e\,\sin(E'))}{\pi ab} \\ 
	&= \dfrac{T}{2\pi}(E'+e\,\sin(E'))
	\label{eqn:t_psi=} \;.
	\end{split}
	\end{equation}
\end{linenomath*} 
Finally, the eccentric anomaly, $E$, is related to the true anomaly,
$\Theta$, by 
\begin{linenomath*}
	\begin{equation}
	E=2\tan^{-1} \left[\sqrt{\dfrac{1-e}{1+e}}\tan \left(\dfrac{\Theta}{2}\right)\right] \;.
	\end{equation}
\end{linenomath*}
${E' = \pi - E}$, ${\Theta = \pi-\psi}$, and here ${\psi = \half \delta}$. Therefore,
\begin{linenomath*}
	\begin{equation}
	E'=\pi - 2\tan^{-1} \left[\sqrt{\dfrac{1-e}{1+e}}\tan \left(\dfrac{\pi-\half \delta}{2}\right)\right] \;.
	\end{equation}
\end{linenomath*}
This, with Eqn.~(\ref{eqn:t_psi=}), gives the time of flight.

\subsection{Altitude Model Fitting}
\label{app:model:altitude}

The ``simulation-fit'' model described in section~\ref{sec:data:altitude} is a sum
of two Chamberlain distributions with three free parameters: 
the relative amplitude of the two distributions and their different temperatures. 
The total amplitude is also free but
is simply set by the observed density at altitude.

In order to fit these parameters at all local times of day and latitudes, 
we first obtained the simulated density at equilibrium for a range of altitudes 
(from 0 to 140~km in 10~km steps). The best-fit parameters for each square degree bin 
were then determined by finding the minimum $\chi^2$ for the simulation data.

These best-fit parameters are publicly available at
\href{http://www.icc.dur.ac.uk/index.php?content=Research/Topics/O13}{icc.dur.ac.uk/index.php?content=Research/Topics/O13}.
In our steady-state model, the argon density
near the end of the lunar night is extremely low, especially at high latitudes and altitudes. 
The LADEE data in these same regions were discarded, 
as discussed in section~\ref{sec:data}, so this was irrelevant for our analysis of the dataset.
Thus, the best-fit parameters in these lowest-density regions were not examined in detail and 
may suffer from the fewer observed particles and high scatter in the simulation results.

This model is dependent on the argon distribution with time of day in the simulation. 
We repeated the analysis with a non-physical isotropic distribution, 
to quantify the maximum possible error that from discrepancies between 
our model and the real distribution. This amounted to $\sim$7\% near the end of the night
and below 1\% elsewhere (within $\pm30^\circ$ latitudes).

\subsection{Squirrelling}
\label{app:model:surface interaction}

We can estimate the statistical effect this downwards migration could
have on the exosphere's distribution with simple, order-of-magnitude
considerations. The regolith temperature just 2~cm below the surface
never drops below 200~K, and by 10~cm only varies by 15~K either side
of 255~K \citep{Teodoro+2015}. 

For particles randomly migrating in the regolith throughout the lunar
night ($\sim{10^6}$~s), with typical steps of size ${\lambda\sim1}$ $\mu$m 
(around and between grains of $\sim\mu$m-mm diameters) and residence times of ${t_{\text{res}}\sim10^{-7}}$~s 
(for ${T=255}$~K and ${Q=28}$~kJ~mol$^{-1}$; 
so negligible traversal times of $\sim$10$^{-9}$~s at thermal speeds),
they will take ${N\sim10^{13}}$~steps, random walking a distance of
${\lambda\sqrt{N}\sim1}$~m.
This implies that a significant number of
particles could bury themselves down into the regolith during the
night, with the dense source of particles residing on the
surface. By symmetry (since the temperatures below the surface are
similar at all times), they will take a similar amount of time 
(the order of half a lunar day) to 
migrate out of the regolith and reenter the exosphere. This suggests
a population of particles that squirrel into the regolith during the
night and typically reenter the exosphere during the day.

%

\acknowledgments

This work was supported by the Science and Technology Facilities Council (STFC) 
grant ST/L00075X/1, and used the DiRAC Data Centric system at Durham University, 
operated by the Institute for Computational Cosmology on behalf of the 
STFC DiRAC HPC Facility (www.dirac.ac.uk). 
This equipment was funded by BIS National E-infrastructure capital grant ST/K00042X/1, 
STFC capital grants ST/H008519/1 and ST/K00087X/1, 
STFC DiRAC Operations grant ST/K003267/1 and Durham University. 
DiRAC is part of the National E-Infrastructure. 
JAK is funded by STFC grant ST/N50404X/1. 
RJM is supported by the Royal Society. 
The LADEE and Diviner data were obtained from the NASA Planetary Data System 
(specifically the ``LRODLR\_1001 PRP'' and ``Derived Data'' bundles respectively). 
We thank Mehdi Benna for helpful comments regarding the processing of the LADEE data 
and instrument background. 
The simulation code is publicly available at
\href{http://www.icc.dur.ac.uk/index.php?content=Research/Topics/O13}{icc.dur.ac.uk/index.php?content=Research/Topics/O13}.
We also thank the two reviewers for their helpful comments.

%
%
%
%
%
%
%
%
%





\end{document}